\documentclass[%
reprint,amsmath,amssymb,aps,prb,superscriptaddress,longbibliography,
floatfix,
]{revtex4-2}

\usepackage{graphicx}% Include figure files
\usepackage{dcolumn}% Align table columns on decimal point
\usepackage{bm}% bold math
\usepackage{epstopdf}
\usepackage{float}
\usepackage{braket}
\usepackage{dsfont}
\usepackage{amsmath}
\usepackage{siunitx} % typesets numbers with units very nicely
\usepackage[italicdiff]{physics}
\usepackage[T1]{fontenc}
\usepackage[dvipsnames]{xcolor}
\usepackage{color,soul}
\usepackage{lipsum}
\usepackage[colorlinks=true, allcolors=blue]{hyperref}

\begin{document}

\title{Characterization of individual charge fluctuators in Si/SiGe quantum dots}

\author{Feiyang Ye}
\thanks{These authors contributed equally.}

\author{Ammar Ellaboudy}
\thanks{These authors contributed equally.}

\affiliation{Department of Physics and Astronomy, University of Rochester, Rochester, NY, 14627 USA}

\author{Dylan Albrecht}

\author{Rohith Vudatha}

\author{N. Tobias Jacobson}
\affiliation{Sandia National Laboratories, Albuquerque, NM, 87185 USA}

\author{John M. Nichol}
\email{john.nichol@rochester.edu}
\affiliation{Department of Physics and Astronomy, University of Rochester, Rochester, NY, 14627 USA}

\begin{abstract}
Electron spins in silicon quantum dots are excellent qubits due to their long coherence times, scalability, and compatibility with advanced semiconductor technology. Although high gate fidelities can be achieved with spin qubits, charge noise in the semiconductor environment still hinders further improvements. Despite the importance of charge noise, key questions about the specific nature of the fluctuators that cause charge noise remain unanswered. Here, we probe individual two-level fluctuators (TLFs) in Si/SiGe quantum dots through simple quantum-dot transport measurements and analyses based on the Allan variance and factorial hidden Markov modeling.
We find that the TLF switching times depend sensitively on gate voltages, decrease with temperature, and depend on the current through a nearby quantum dot. A model for the data of the primary TLF we study indicates that it may be a bistable charge dipole near the plunger gate electrode, heated by current through the sensor dot, and experiencing state transitions driven not by direct electron-phonon coupling but through some other mechanism such as coupling to electrons passing through the sensor dot. 
\end{abstract}

%\date{\today}\\

\maketitle

\section{Introduction}
Electron spins in Si quantum dots have long coherence times, enabling the high-fidelity gate operations essential for quantum computation~\cite{burkard2023semiconductor}.
%\cite{veldhorst2015two, watson2018programmable, mills2022two, noiri2022fast, mkadzik2022precision, weinstein2023universal}.
Although isotopically-purified Si spin qubits experience minimal magnetic noise, charge noise in Si quantum dots is a major obstacle limiting gate fidelities \cite{kranz2020exploiting}.
Most gate operations for quantum-dot spin qubits depend on the precise control of local electrostatic potentials \cite{burkard2023semiconductor}, and charge noise causes unwanted fluctuations of these potentials, reducing gate fidelities.

In semiconductor quantum dots, charge noise typically has a $1/f$-like power spectrum, and is thought to be linked to the presence of two-level fluctuators (TLFs) in the semiconductor~\cite{Paladino2014,burkard2023semiconductor,Freeman2016,Yoneda2018,mi_landau-zener_2018,connors2019low,struck2020low,petit_spin_2018,rudolph2019long,kranz2020exploiting,elsayed2022low,holman20213d,connors2022charge,paquelet2023reducing}. Previous work on charge-noise spectral densities has indicated that charge noise levels are above the expected levels from electronics (see, e.g, Ref.~\cite{connors2022charge}), and that the noise sources are located within the device, perhaps near the semiconductor surface or near the quantum well ~\cite{Freeman2016,petit_spin_2018,connors2019low,kranz2020exploiting,connors2022charge,paquelet2023reducing}.
Although TLFs and two-level systems have been studied intensely in other physical systems such as glasses~\cite{anderson1972anomalous,phillips1987two}, transistors \cite{kirton1989noise}, and other qubit platforms~\cite{muller2019towards}, whether or not TLFs exist in Si quantum dots and what their properties are remain open questions. 
Due to the close relationship between charge noise and gate fidelities in spin qubits,
understanding the origin of electrical noise in Si quantum dots is critically important.
%Experiments probing individual two-level charge fluctuators are crucial to suggest methods to mitigate or even remove the TLFs in semiconductor spin qubits.

In this work, we find that TLFs indeed cause electrical noise in Si/SiGe quantum dots. Moreover, by analyzing our data with the Allan variance and factorial hidden Markov modeling, we can determine the properties of individual TLFs. We demonstrate that the TLFs are extremely sensitive to gate voltages. We also find that the TLF switching times decrease with temperature, and we find that the temperature of the TLFs depends on current through the sensor quantum dots. By comparing our data for one TLF to different models, we exclude scenarios involving pure phonon-assisted tunneling or pure thermal activation. Instead, we can explain our data using a model that incorporates switching resulting from both thermal activation and tunneling mediated not by electron-phonon coupling but by a different mechanism, such as coupling to electrons in the two-dimensional electron gas. %by a bosonic bath. 

This work confirms the relevance of TLFs for electrical noise in Si quantum dots and establishes the viability of characterizing the properties of individual TLFs. In the future, this ability may enable the determination of the exact nature of the TLFs and the development of methods to eliminate charge noise.

\section{Experimental setup}

\begin{figure}[t]
\centering
{\includegraphics[width=0.5 \textwidth]{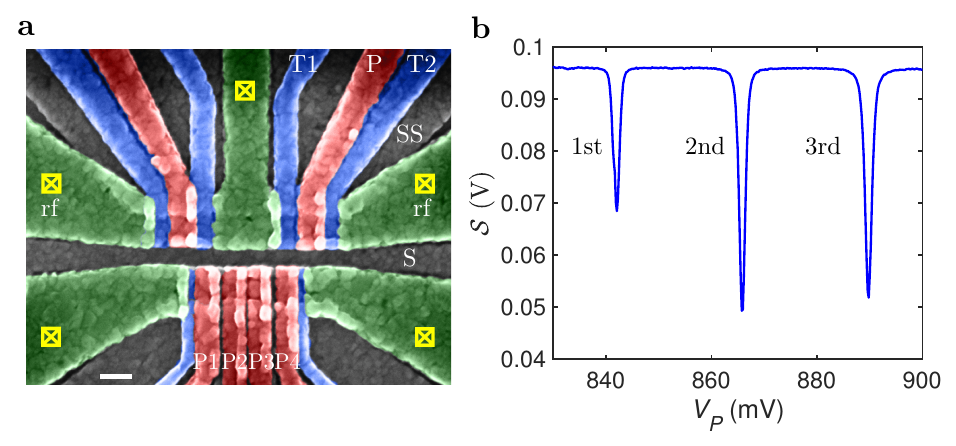}}
\caption{
\textbf{Measurement setup.}
\textbf{a} False-color scanning electron micrograph of a Si quadruple-quantum-dot device nominally identical to the one used here. Plunger gates, tunneling gates, screening gates, and accumulation gates are color-coded as red, blue, gray, and green, respectively. 
The two sensor quantum dots share a drain, and we send the reflectometry signal through ohmic contacts labeled by yellow boxes to the quantum well for rf reflectometry.
The white scale bar is $100~\si{nm}$.
\textbf{b} Sensor-dot conductance peaks measured by rf reflectometry. The transport peaks appear as dips in the reflectometry signal $\mathcal{S}$ because of the phase shift of our reflectometry circuit. 
}
\label{fig:setup}
\end{figure}

The primary device we use (``Device 1'') is a quadruple-quantum-dot device with two rf-equipped charge sensors~\cite{connors2020rapid} fabricated with an overlapping gate architecture (Figure~\ref{fig:setup}a) on an undoped Si/SiGe heterostructure with an 8-nm-thick natural Si quantum well approximately $50~\si{nm}$ beneath the surface. %and a $2~\si{nm}$ silicon cap layer.
The device is cooled in a dilution refrigerator with a base temperature around $10~\si{mK}$. We have also characterized a second device (``Device 2''), a double quantum dot with one rf charge sensor. Device 2 is fabricated on a different, but nominally similar, wafer to that used for Device 1. Device 2 is also measured in a different, but nominally similar, dilution refrigerator to the one used for Device 1. Both devices have a 15-nm-thick aluminum oxide gate dielectric deposited via atomic layer deposition. See the Supplemental Material \cite{SMnote} for more information about the device fabrication.  

We tune the right sensor quantum dot of Device 1 in the Coulomb blockade regime as shown in Fig.~\ref{fig:setup}b, and we set the plunger gate voltage $V_{P}$ on the flank of the transport peak, such that conductance fluctuations of the sensor dot reflect electrochemical potential fluctuations \cite{Freeman2016, connors2019low, paquelet2023reducing}. Of the four quantum dots on the main side, dots 1 and 2 are tuned to accumulation, and dots 3 and 4 are tuned to the $(0,1)$ charge configuration, with $0$ $(1)$ electron in dot $3$ $(4)$. We expect that this tuning suppresses electrical fluctuations in the main side and does not strongly affect the sensor side. We measure the sensor conductance via rf reflectometry \cite{connors2020rapid, 10.1063/5.0088229} and sample the resulting downconverted signal $\mathcal{S}$ at a $60~\si{Hz}$ rate to minimize electronic noise. 

We convert variations in the reflectometry signal $\delta \mathcal{S}$ resulting from charge noise into electrochemical potential fluctuation $\delta \epsilon$ via the equation $\delta \epsilon = \alpha \delta \mathcal{S}/(d\mathcal{S}/dV_{P})$ \cite{connors2019low}. We extract the sensor sensitivity $d\mathcal{S}/dV_{P}$ by differentiating the peak shape at the configuration used for measurements. We find the lever arm $\alpha$ by raising the temperature of the mixing chamber to 500 mK and fitting the peak shape to the predicted Coulomb blockade lineshape, assuming pure thermal broadening. (See Supplemental Figure S1 for further details on the lever-arm measurement as well as electron temperature measurements \cite{SMnote}.)

\section{TLF characterization}
\subsection{Time domain}
\begin{figure*}[t]
\centering
{\includegraphics[width=1 \textwidth]{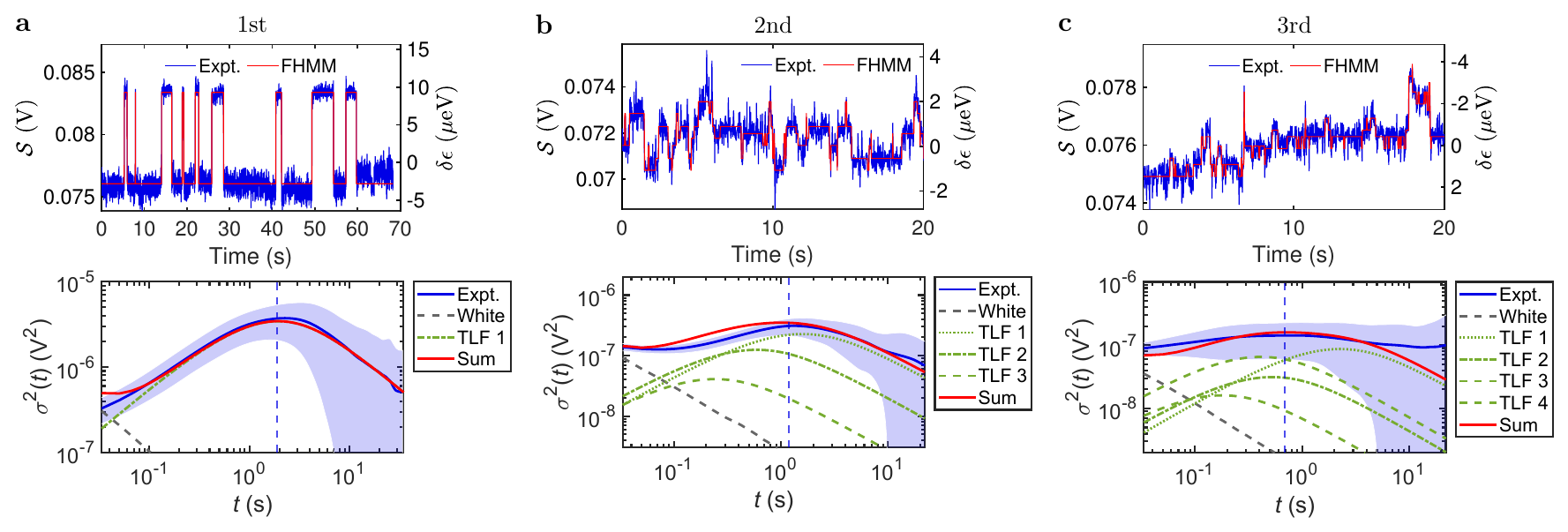}}
\caption{
\textbf{Voltage-dependent noise environment.}
\textbf{a-c} Examples of reflectometry time traces and electrochemical potential fluctuations (top) and Allan variances (bottom) measured from the different transport peaks in Fig.~\ref{fig:setup}b. 
Top: Experimental time series from different transport peaks (blue) and the most likely fluctuator noise trajectory without a white noise component, as estimated from the FHMM using the Viterbi algorithm (red).
Bottom: Experimental Allan variance (blue) and theoretical Allan variances for the predicted TLFs extracted from the FHMM modeling. The blue region is the Allan variance for the experimental data, with the confidence regions estimated from the spread of the 30 repetitions. The vertical  dashed line represents the maximum ($t_{\rm max}$) in the experimental data Allan variance. The green dashed lines represent the analytic Allan variance for the TLFs present in the corresponding time series, as extracted via the FHMM fits, with the addition of the model white noise which goes like $1/t$ (gray). The solid red curve is the sum of the constituent analytical Allan variances.
}
\label{fig:avar}
\end{figure*}

Figure~\ref{fig:avar} shows representative time traces measured on different transport peaks, where we plot both reflectometry signals and electrochemical potential fluctuations. Each time trace clearly shows the presence of one or more random telegraph signals with characteristic electrochemical potential amplitudes of a few to about $10~\si{\mu eV}$.

Interestingly, the character of the noise varies significantly between the different transport peaks. The data from the first transport peak show random telegraph noise with a constant step height, indicating a single dominant TLF (Fig.~\ref{fig:avar}a). However, the data from the second (Fig.~\ref{fig:avar}b) and third (Fig.~\ref{fig:avar}c) transport peaks show multiple step heights, suggesting the presence of several TLFs. As these data imply, and as we will discuss further below, the characteristic switching time of the TLFs depends sensitively on gate voltages, explaining the discrepancy between transport peaks. In particular, the data we discuss below suggests that the TLF on the first transport peak is frozen out on the second and third transport peaks, revealing the other TLFs.

Before proceeding, we note that most previous experiments on charge noise in Si quantum dots have reported power spectral densities, instead of time-domain measurements. To link the time-domain results we discuss below to previous frequency-domain measurements, including qubit measurements, we note that Device 2 in this work is the same device described in Ref.~\cite{connors2022charge}. In that earlier work, which took place in a different cryostat, we found that the qubit associated with Device 2 had an approximately $1/f$-like charge-noise power spectrum, with a relatively low power spectral density at 1~Hz of 0.42~$\mu$eV$^2$/Hz. Despite the $1/f$-like spectrum, early time-domain qubit measurements on that device revealed the presence of discrete electrical fluctuators (see Fig. 3b of Ref.~\cite{connors2022charge}). We emphasize that the presence of resolvable TLFs in a Si quantum dot is still compatible with relatively low noise levels and the presence of approximately $1/f$-like noise spectra. 

\subsection{Allan variance}
In general, extracting the properties of individual TLFs within an ensemble poses a challenge. Indeed, most previous measurements of charge noise in semiconductor quantum dots have used analyses based on the power spectral density. However, disentangling the effects of individual TLFs from a power spectrum is challenging because relatively few TLFs are required to produce an approximately smooth, featureless $1/f$-like spectrum. 

To overcome this challenge and to analyze individual 
TLFs, we analyze our time series data using two methods. The first of these methods involves the Allan variance \cite{Allan1966, VANVLIET1982261, PRINCIPATO200775}. Roughly speaking, the Allan variance quantifies how much a signal changes on average after a time lag $t$. In fact, the Allan variance of the random telegraph signal from an individual TLF has a peak at a time lag nearly equal to the average switching time of the TLF \cite{PRINCIPATO200775, burnett2019decoherence, yang2023locating}.  Specifically, if $\Gamma_{01}$ and $\Gamma_{10}$ are the forward and reverse transition rates of a single TLF, respectively, the peak of the Allan variance $\sigma^2(t)$ of the random telegraph noise from that TLF occurs at $t_\text{max} \approx 1.893/(\Gamma_{01} + \Gamma_{10}) = 0.946/\gamma$, where the average switching rate $\gamma = (\Gamma_{01} + \Gamma_{10})/2$~\cite{VANVLIET1982261, PRINCIPATO200775}.
The Allan variance has also been used previously to analyze the contribution of individual fluctuators to the relaxation rate of superconducting qubits \cite{burnett2019decoherence,yang2023locating}.

Examples of Allan variances $\sigma^2(t)$ as a function of time lags $t$ are plotted in Fig.~\ref{fig:avar}. In Fig.~\ref{fig:avar}a, there is a pronounced single peak in the Allan variance, reflecting the presence of the dominant TLF on the first transport peak. In Fig.~\ref{fig:avar}b, the Allan-variance peak is less pronounced and has shifted to shorter times, reflecting the presence of additional, faster TLFs on the second transport peak. In Fig.~\ref{fig:avar}c, corresponding to the third transport peak, the Allan-variance peak is even less pronounced.

\subsection{Factorial hidden Markov modeling}
The second method we use to analyze our data is a factorial hidden Markov model (FHMM) \cite{ghahramani1995factorial,Albrecht2023,puglisi2017random}, which has been used extensively to model the hidden states that contribute to a signal, including in the study of charge traps in other microelectronic devices~\cite{puglisi2015complete}.  The FHMM directly models the time series as a combination of $d$ independent, hidden, $k$-level fluctuators.  For our purposes we fix $k=2$, focusing on an ensemble of independent TLFs. These fluctuators are combined together with individual weights, modeling the output as a multivariate normal distribution:
\begin{equation}
\vec{y}_{t} \sim \mathcal{N}(W \cdot s_{t}, C),
\end{equation}
where $W$ are the weights, $s_t$ are the fluctuators, represented as an indicator vector, and $C$ is the covariance.  The model also requires transition matrices for all of the fluctuators as additional parameters.  We use exact Expectation-Maximization (EM) to fit the model parameters.

To illustrate the capabilities of the FHMM approach, we fit the data from the different transport peaks. Figure~\ref{fig:avar} shows representative time traces from the different transport peaks, as well as the Viterbi (most probable) path for the hidden fluctuators' trajectories. Based on a cross-validating model selection procedure (see Supplemental Material \cite{SMnote}), we fit $d=1$, $d=3$, and $d=4$ models to the 1st, 2nd, and 3rd datasets, respectively. In Fig.~\ref{fig:avar}, we plot the Allan variance of the experimental data as compared to the Allan variance of data generated by the best-fit models. The averages match quite well, mostly lying in the confidence bands. Also plotted is the deconstruction of the signal into the Allan variances of the individual TLF components, and a white noise background. This showcases the blind source separation capability of the FHMM technique, separating the noise time series into independent signal components, where these components sum to match the experimental noise time series and the Allan variance.

%Both methods we use in this work--the Allan variance and FHMM technique--enable characterizing individual TLFs when multiple fluctuators are present. 
As discussed in this section, the dynamics of a single dominant TLF are conveniently characterized by evaluating the Allan variance and distribution of TLF occupations, while in the regime of multiple significant TLFs the FHMM approach provides a powerful means of identifying the dynamics of each constituent. 
As we will discuss next, for example, these capabilities enable us to understand why and how charge noise changes between transport peaks. In the remainder of this work, we will leverage this ability to extract detailed information about individual TLFs. Altogether, these results illustrate how studying individual TLFs can complement traditional ensemble measurements to improve understanding of charge noise. %  In addition to providing detailed modeling, the extraction of physically meaningful parameters, and uncertainty quantification of the TLF content at a particular point in voltage space, the FHMM modeling provides a quantitative method to analyze the variability of the TLF content when changing transport peaks, which is seen here to be drastically different from one peak to another.

\section{Voltage dependence}
\subsection{Observations}
\begin{figure*}[!ht]
\centering
{\includegraphics[width=1 \textwidth]{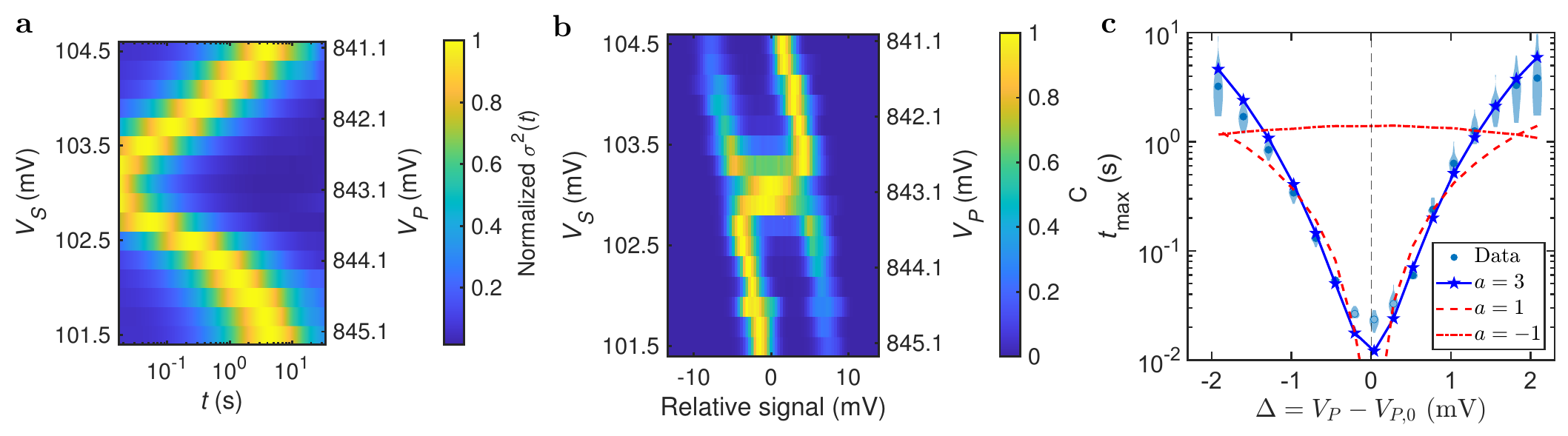}}
\caption{
\textbf{Voltage dependence of TLF switching rate and bias.}
% \textbf{a} Examples of the time traces at different voltage configurations. Both switching time and occupation bias are very sensitive to gate voltage.
\textbf{a} Normalized Allan variance as a function of time and gate voltage. The left (right) axis plots the screening (plunger) gate voltages.
The peak of the Allan variance is approximately at the average TLF switching time.
\textbf{b} Histograms of the time traces showing the occupation of the TLF states at different gate voltages. The most likely state changes on either side of $V_{P,0}$. At each gate voltage, the count C is normalized by its maximum value.
\textbf{c} Allan variance peak observation time $t_{\mathrm{max}} \approx 1.893/(\Gamma_{01} + \Gamma_{10})$ as a function of voltage offset from the zero-bias point $V_{P,0} \approx 843.1 \ \mathrm{mV}$. Plotted in blue is the measured Allan variance, with points and violin plots denoting medians and distributions of Allan variance across 30 repeats, respectively. The observed mean switching time $\tau$ exhibits a dependence $\vert \Delta E \vert^{a} \tanh(\Delta E / 2 k_{B}T)$ with $a \approx 3$, while a dependence with $a$=$1$~($-1$) is expected from coupling to electrons (phonons) \cite{phillips1987two,Paladino2014}.   The open data points are excluded from the fit due to aliasing artifacts arising from transition rates approaching or exceeding the 60 Hz sample rate of the measurement (see Supplemental Material \cite{SMnote}).
}
\label{fig:comp}
\end{figure*}

As depicted in Fig.~\ref{fig:avar}, the switching times and amplitudes of the observed TLFs depend on which conductance peak is used to measure charge noise, suggesting that the TLFs are sensitive to gate voltages.
To explore this in detail, we tune the right sensor dot of Device 1 to the first transport peak, where the time traces feature a prominent single TLF (Fig.~\ref{fig:avar}a). (The data displayed in the remainder of the main text will focus on this TLF. See the Supplemental Material \cite{SMnote} for data on other TLFs in Device 1 and Device 2.) To study the voltage dependence of this TLF, and to eliminate effects associated with changing the sensor quantum dot chemical potential, and thus the charge-noise sensitivity, we keep the quantum dot chemical potential fixed by sweeping the plunger and screening gate voltages together.  For each voltage configuration, we measure a 34-minute-long time series and compute the normalized Allan variance, which we define as the Allan variance divided by its maximum value, and the results are plotted in Fig.~\ref{fig:comp}a.
% Figures~\ref{fig:comp}a-c plot examples of the time-traces at different voltages. In Fig.~\ref{fig:comp}a the TLF switches slowly, and the high-signal state is occupied more frequently.  In Fig.~\ref{fig:comp}b, the TLF switching time greatly decreases and the two states are almost equally occupied. In Fig.~\ref{fig:comp}c, the TLF switches slowly again, but the low-signal state is more frequently occupied.

The data show that the TLF switching time depends strongly on the gate voltages, reaching a minimum value at a particular voltage $V_{P,0}$ and then increasing again away from $V_{P,0}$. Figure~\ref{fig:comp}b reveals further information about the most likely state of the TLF as a function of gate voltages. At each gate voltage configuration, we plot the histogram of the sensor time traces to show the occupation of the TLF states.
We subtract the mean signal at each voltage to facilitate a comparison between different voltages. At plunger gate voltages larger than $V_{P,0}$, the TLF occupies the $0$ state (lower relative signal) more often, while at lower voltages, the TLF occupies the $1$ state (higher relative signal) more often. Supplemental Figures S2 and S5 show data from other voltage-dependent TLFs in both Device 1 and Device 2 \cite{SMnote}.

\subsection{Interpretation}

\subsubsection{Explanation for the variation in noise between transport peaks}
This sensitive voltage dependence is a likely explanation for the variation of charge noise between the transport peaks in Fig.~\ref{fig:avar}. Considering that the switching time of the TLF changes by several orders of magnitude as the gate voltages change by only a few mV, and considering that neighboring transport peaks are tens of mV away from each other, the single TLF observed on the first transport peak is likely frozen in one of its states on the other transport peaks. Multiple groups have reported variations in charge noise between transport peaks~\cite{connors2019low,zwerver_qubits_2021,paquelet2023reducing}, and the sensitive voltage dependence we observe is a possible explanation for this variation.

\subsubsection{Two-state system model}
If we assume that the TLF is in thermal equilibrium with a reservoir at temperature $T$, the occupation ``bias'' $\mathcal{B}$ between the $0$ and $1$ states of the TLF obeys
\begin{equation}
%\mathcal{B} \equiv N_L/N_R = \exp\qty(-\Delta E/k_B T),
\mathcal{B} \equiv N_0/N_1 = \exp\qty(-\Delta E/k_B T),
\label{eq:bias}
\end{equation}
where $i=0,1$ indexes the state, $N_i$ denotes the population of the state, and $k_B$ is Boltzmann's constant. In the high-temperature regime, thermal activation causes the TLF to switch, and Eq.~\eqref{eq:bias} follows from the principle of detailed balance. In this case, $\Delta E$ is the difference in energy, or asymmetry, between the two potential-energy minima representing the two classical states. In the low-temperature limit, environment-assisted tunneling drives transitions between two quantum states, and Eq.~\eqref{eq:bias} follows from the Boltzmann factor. In this case, $\Delta E$ represents the energy difference, or detuning, between the quantum states. In either of these regimes, the data of Fig.~\ref{fig:comp}b are consistent with a scenario in which the energy difference between the two states decreases and then changes sign as the voltage is changed through $V_{P,0}$. One plausible hypothesis to explain this behavior is that the gate voltages alter the detuning or asymmetry of the TLF, with the minimum switching time configuration corresponding to a near-degeneracy between the two states. Our results also suggest that at least for this TLF, the switching time increases with the magnitude of the detuning or asymmetry. This fact has significant implications for the precise nature of this defect and to which thermal reservoirs it is coupled, as we discuss next.

\subsubsection{Comparison with microscopic models}
In the high-temperature limit, a common model assumes that the TLF is thermally activated such that the average switching time, defined as the inverse of the average switching rates $2/\qty(\Gamma_{01} + \Gamma_{10})$, $\tau(E_{b},\Delta E, T) \propto e^{E_{b}/k_{B}T}/\cosh \qty(\Delta E/2 k_B T )$ \cite{dutta1981low, kirton1989noise}, where $E_{b}$ is an activation barrier energy. Our data, which show that the switching time increases with the energy bias, are inconsistent with a pure thermal-activation model.  
  
In the low-temperature limit, where reservoir-assisted tunneling dominates, two common models involve two-level tunneling states coupled to phonon or electron reservoirs. In these cases, the average switching time is given by $\tau(\Delta E,T) \propto \Delta E^{a} \tanh(\Delta E/2 k_B T)$, where the exponent $a$ may be equal to $+1$ ($-1$) for the interaction with electrons (phonons) \cite{phillips1987two,Paladino2014} and $\Delta E$ is the energy difference between two states.
For the case of a tunneling system interacting with a phonon reservoir (e.g. $a=-1$), the switching time decreases with the energy splitting. Depending on phonon characteristics and geometry, the phonon-induced relaxation time may decay even faster as a function of energy difference, with $a<-1$ \cite{Fedichkin2004}. 

Our data are also inconsistent with such a phonon-mediated tunneling model (Fig.~\ref{fig:comp}c), because the switching time increases with the bias. A tempting explanation for charge noise in quantum dots involves double-well-potential defects in gate oxide layers that switch between two states via pure phonon-assisted tunneling, but this picture is not consistent with our data for this TLF. Instead, the qualitative features in our data are consistent with electron-assisted tunneling, which predicts an increased switching time with increasing detuning.

\section{Temperature dependence}
\subsection{Observations}
\begin{figure*}[ht!]
\centering
{\includegraphics[width=1 \textwidth]{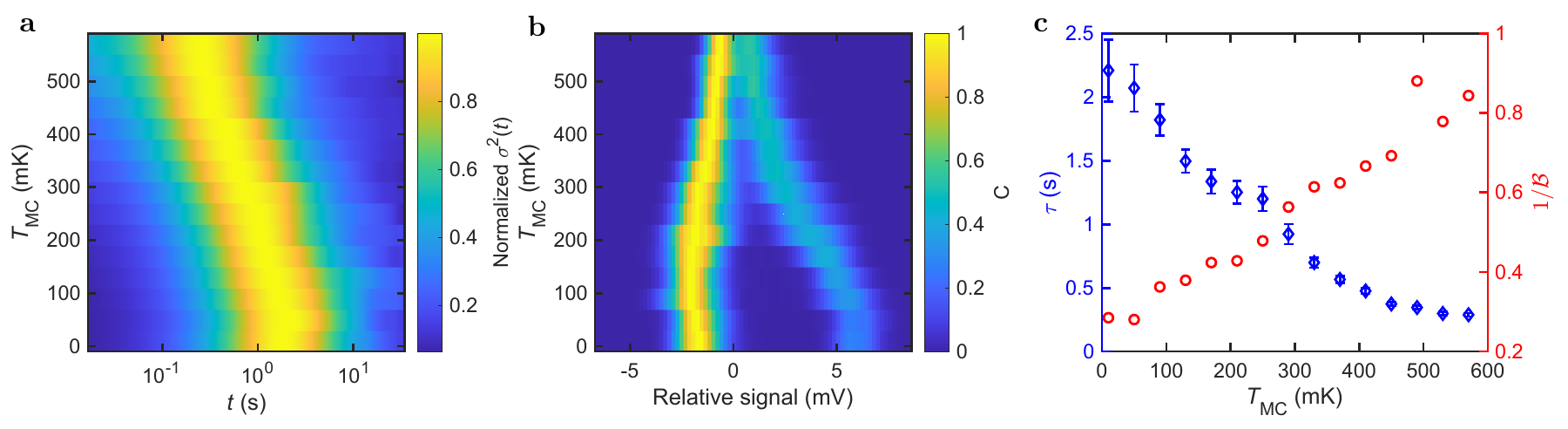}}
\caption{
\textbf{Temperature dependence.}
\textbf{a} Normalized Allan variance as a function of time and mixing chamber temperature.
\textbf{b} Histograms of the time traces showing the occupation at different mixing chamber temperatures. The mean signal is subtracted at each temperature. The signal difference between the two states decreases at elevated temperature because of thermal broadening of the transport peak.
\textbf{c} Left: Switching time as a function of mixing chamber temperature.
Right: The occupation bias between the excited state and the ground state versus mixing chamber temperature.
At elevated temperatures, the TLF switching time tends to decrease rapidly and the excited state is occupied more frequently.  We extract the bias by fitting histograms of our data to the sum of two Gaussians (see Supplemental Material \cite{SMnote}).
% \textbf{b} Example of time-trace when the mixing chamber temperature is at $10~\si{mK}$.
% \textbf{c} Example of time-trace when the mixing chamber temperature is at $570~\si{mK}$.
}
\label{fig:MCtemp}
\end{figure*}

As discussed above, we observe that the TLF switching times depend on gate voltages. We now show that they depend on temperature. To study the effect of temperature variations on the TLF switching times, we sweep the mixing chamber temperature $T_\text{MC}$ from $10$ to $570$ $\si{mK}$ with an increment of $40~\si{mK}$  and measure the switching time at each temperature (Fig.~\ref{fig:MCtemp}a). During the entire temperature sweep, all gate voltages are fixed. As expected, the switching time decreases with temperature, in agreement with our expectation that the switching is driven by thermal fluctuations.
Figure~\ref{fig:MCtemp}b plots histograms of the time traces to reveal the occupation of the TLF states at each temperature. At high temperature, the TLF occupies the excited state (higher relative signal) more frequently.
The bias between the excited state and ground state $1/\mathcal{B}$ increases monotonically with temperature (Fig.~\ref{fig:MCtemp}c), as expected.
%Figures~\ref{fig:MCtemp}b and c show time traces at different temperatures and illustrate the change in switching time with temperature.  
Supplemental Figure S3 shows a similar dependence of the switching time on temperature in other TLFs in Device 1 \cite{SMnote}.

\subsection{Interpretation}
\subsubsection{Comparison with microscopic models} 
As shown later in Fig.~\ref{fig:FullFit}b, both the forward and reverse switching times for this TLF, as extracted via FHMM, decrease with temperature. This behavior is inconsistent with a simple charge-trap model involving elastic tunneling to and from an electron reservoir. In such a model, the forward and reverse switching times depend oppositely on temperature~\cite{Beaudoin2015}. 

\subsubsection{Thermalization mechanism} 
In principle, a careful measurement of the TLF switching time vs. temperature should help to determine the microscopic nature of the TLF, or at least to what thermal reservoirs it is coupled. However, assessing the quantitative agreement between these data and different TLF models is challenging for a variety of reasons. For example, we observe that the switching time appears to decrease less rapidly at very low temperatures $T_\text{MC} \le 50~\si{mK}$. A number of effects could result in this behavior, such as environment-assisted tunneling or poor thermalization of the device to the mixing chamber. However, as discussed in more detail in the next section, there is another effect that complicates this analysis. 
We find that the current through the sensor quantum dot, which we use to measure the TLF, appears to heat the TLF. Thus, at the lowest mixing chamber temperatures, the effective temperature of the TLF is higher than our thermometer reading, because of this excess heating. Moreover, this effect cannot be explained through an elevated electron temperature in the usual sense. We discuss this behavior in more detail in the following sections.   

\section{Local heating}
\subsection{Observations}
\begin{figure*}[ht!]
\centering
{\includegraphics[width=1 \textwidth]{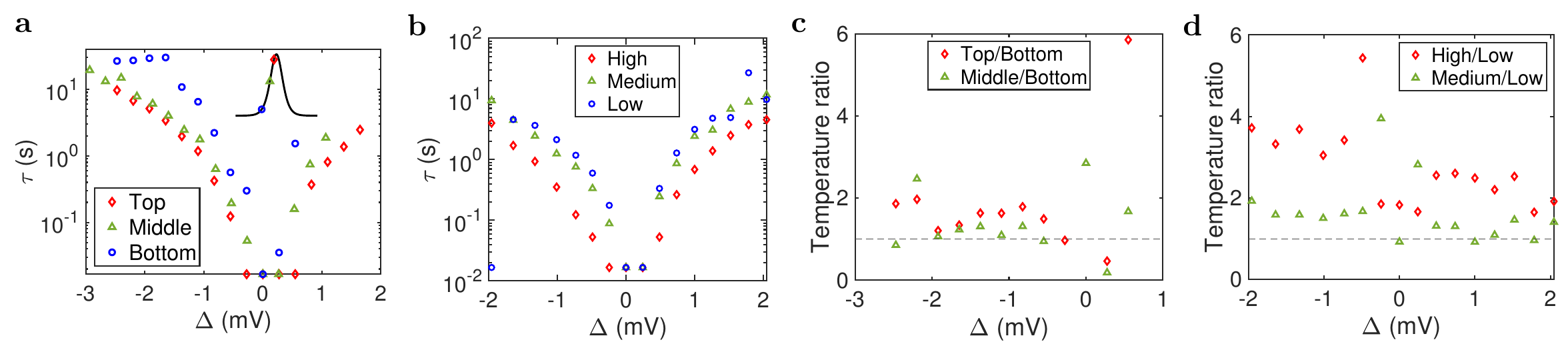}}
\caption{
\textbf{Local heating.}
% \textbf{a} Example time-trace with high dot conductance.
% \textbf{b} Example time-trace with low dot conductance. In \textbf{a} and \textbf{b}, the TLF detuning $\Delta$ is fixed.
\textbf{a} Switching time vs. gate voltage at different dot conductances.
The switching times decrease with increasing conductance.
We omit the switching-time data near the bottom of the conductance peak when the TLF switching time is comparable to or longer than our data acquisition time.
Inset: We set the plunger gate voltage to different positions on the transport peak to vary the conductance.
\textbf{b} Switching time vs. gate voltage at different excitation powers. The switching times decrease with increasing power.
% \textbf{e} Example time-trace with high rf power when $\Delta = -1~\si{mV}$.
% \textbf{f} Example time-trace with low rf power when $\Delta = -1~\si{mV}$.
\textbf{c} Effective temperature ratio vs. gate voltage between different sensor-dot conductances.
\textbf{d} Effective temperature ratio vs. gate voltage between different sensor-dot excitation powers.
In \textbf{c} and \textbf{d}, the TLF temperature increases with more sensor-dot current. The horizontal line indicates equal temperature.
}
\label{fig:localheating}
\end{figure*}

Surprisingly, we observe that the TLF switching times decrease with increasing conductance of the sensor dot. Figure~\ref{fig:localheating}a shows the results of experiments similar to those of Fig.~\ref{fig:comp}a, except that we fix the chemical potential of the sensor dot at different levels for the three different scans. To compensate for the shift in the TLF detuning for the different cases, the $x$-axis in all cases is $\Delta=V_P-V_{P,0}$. Figure~\ref{fig:localheating}a plots the extracted switching times for the different conductances, and we observe that the switching time for the same value of $\Delta$ decreases with the conductance. We have also verified that the same behavior holds on the other side of the transport peak.  

We also measure the TLF while varying the source-drain bias voltage by adjusting the sensor rf excitation, at fixed sensor dot conductance.
Figure~\ref{fig:localheating}b shows the gate-voltage dependence of the TLF switching time with different rf powers. We find that switching times decrease with increasing rf power.
%, as shown by time traces in Figs.~\ref{fig:localheating}e,f.

Raising the mixing chamber temperature $T_\text{MC}$ to $200~\si{mK}$ causes these effects to become less pronounced (Supplemental Figure S7 \cite{SMnote}). Supplemental Figure S4 shows similar changes to the switching time of other TLFs in Device 1 and Device 2 with dot conductance \cite{SMnote}.

\subsection{Interpretation}

\subsubsection{TLF heating}
These data show that the TLF switching time decreases with current through the sensor dot.
If we assume that the TLF energy difference or asymmetry does not depend on the dot conductance or rf power, we can calculate a relative temperature change using the relation $T \propto 1/|\ln \mathcal{B}|$ (Figs.~\ref{fig:localheating}c,d).
Omitting the bias data near $V_{P,0}$ because the histogram peaks are not well separated, and using the data near the bottom of the transport peak as a temperature base, we compute that the effective temperature ratio between the top (middle) and bottom of the transport peak is $1.61 \pm 0.09$ ($1.28 \pm 0.18$) (Fig.~\ref{fig:localheating}c).
Likewise, Figure~\ref{fig:localheating}d shows that the effective temperature ratio between high (medium) and low rf power is calculated as $2.76 \pm 0.20$ ($1.39 \pm 0.08$).

\subsubsection{Origin of the heating} 
The origin of this heating effect is not clear. It could be related to Joule heating in the electron reservoirs or the non-equilibrium distribution of occupied states during current flow. We also note that a similar type of charge sensor-induced TLF heating effect has been observed previously in other systems \cite{Kenyon2001,Gustafsson2013}. Future studies on the origins of the local heating effect may help to understand the precise location and nature of the TLFs. 

\subsubsection{Effect of electron temperature} 
Our data show that the electron temperature, as it is most commonly measured, does not by itself determine TLF switching rates. Electron temperatures are frequently assumed not to depend on gate voltages. However, our data show that TLF temperatures can change drastically with a small change in gate voltages even at what one would normally consider a fixed electron temperature. It could be that the electron temperature varies over the transport peak~\cite{kautz1993self}, but quantifying this effect will require further experiments and modeling to determine the thermal properties of the electron reservoirs in our system.  

\section{Gate-voltage sensitivity}
\subsection{Observations}
\begin{figure*}[!ht]
\centering
{\includegraphics[width=1 \textwidth,trim=0 0 0 0,clip]{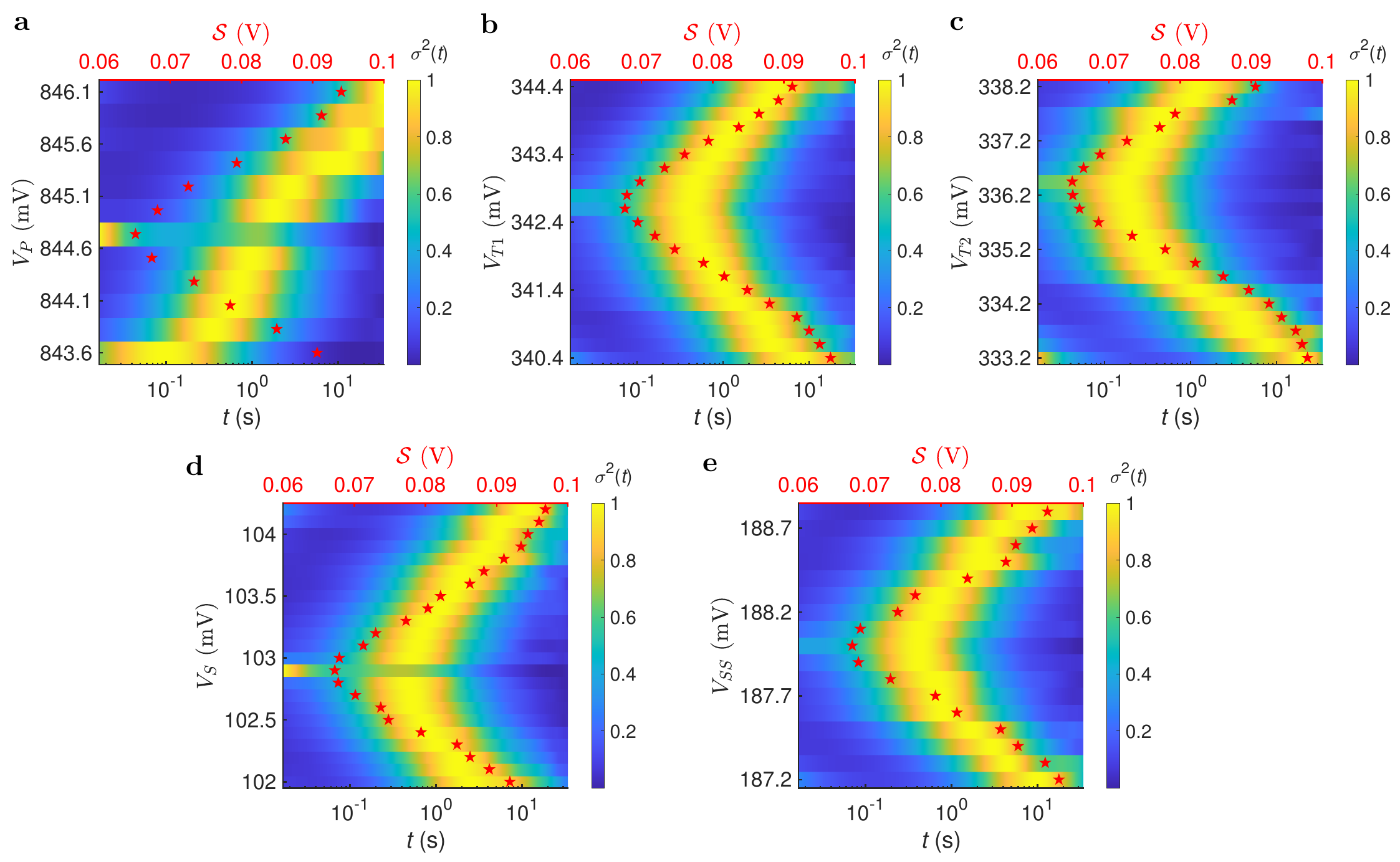}}
\caption{
\textbf{Gate-voltage sensitivity.}
\textbf{a}-\textbf{e} The voltage response of the TLF switching times to different sensor gate voltages. We perform the measurements here at $T_\text{MC}=200~\si{mK}$. In red stars, we plot the mean reflectometry signals $\mathcal{S}$ vs. gate voltages to show the transport peak. At the top of the transport peak, the sensor loses its sensitivity to electrochemical potential fluctuations, and the Allan variance does not reflect charge noise. The TLF switching time responds symmetrically to most gates, except the plunger gate. The monotonic change of the switching times across the conductance peak suggests that the TLF detuning effect induced by the plunger gate is stronger than the local heating effect.
}
\label{fig:gate}
\end{figure*}

In Si/SiGe quantum dots, the locations of the TLFs are not known. In principle, the TLFs may arise from defects in multiple locations \cite{connors2019low,paquelet2023reducing} including the disordered dielectric layer near the semiconductor surface, the $\text{SiGe}$ buffer layer, the buffer/quantum well interface, and the buried $\text{Si}$ quantum well.
To gain insight into the location of the fluctuator, we measure how the TLF switching time changes in response to different gate voltages (Fig.~\ref{fig:gate}). In principle, the most straightforward way to achieve this goal would be to measure how the TLF switching time changes in response to small changes to each of the gate voltages. However, the local heating effect discussed above complicates this procedure. Changes in the heating effect can be eliminated by fixing the sensor dot chemical potential, but this extra constraint means that the gate voltages cannot be changed independently.  

In spite of this difficulty, we can qualitatively measure the response of the TLF switching time to different gate voltages by sweeping the voltages independently such that the chemical potential of the sensor dot traverses the whole transport peak. In this situation, the TLF switching time is affected by both the gate voltage-controlled energy splitting (Fig.~\ref{fig:comp}) and the local heating effect (Fig.~\ref{fig:localheating}).

\subsection{Interpretation}
If the TLF responds primarily to the local heating effect, the switching time should vary symmetrically about the center of the transport peak. However, if the TLF responds primarily to the voltage-induced detuning or asymmetry change, the switching time should vary monotonically across the transport peak.

Figure~\ref{fig:gate} illustrates how the switching time of the TLF changes with different gate voltages. For most gates except the plunger gate, the TLF switching time varies symmetrically across the transport peak. In these cases, we attribute this behavior to the local heating effect described earlier, where the switching time decreases with the conductance of the sensor dot. Under this assumption, the symmetric variation of the switching time indicates weak coupling to the gate electrode. This behavior is consistent with other observations of a quantum-dot-current-dependent TLF heating effect \cite{Kenyon2001}.
% The screening and tunneling gates barely change the splitting energy of the TLF, so that they couple weakly to the TLF.
% Varying these gate voltages changes the dot conductance at a give plunger gate voltage as depicted in Fig.~\ref{fig:gate}a, leading to the local heating effect.
% However, the monotonously increasing switching time verse the plunger gate voltage shown in Fig.~\ref{fig:gate}d indicates that the plunger gate affects largely on the TLF energy, at least compared with other sensor gates.
% The strong coupling between the plunger gate and the TLF can cancel out the local heating effect on the switching time.
In Fig.~\ref{fig:gate}a, however, we observe a monotonic change of the switching time across the center of the transport peak when we sweep the plunger gate, suggesting that the plunger gate couples strongly to the TLF. Supplemental Figure S5 shows the results of similar experiments for other TLFs in Device 1 \cite{SMnote}, and we usually find that one or two gates appear to couple most strongly to the TLFs.

\section{Phenomenological Model}
\subsection{Combined tunneling/activation model}

\begin{figure*}[ht!]
\centering
  \includegraphics[width=0.95 \textwidth,trim=0 0 0 0,clip]{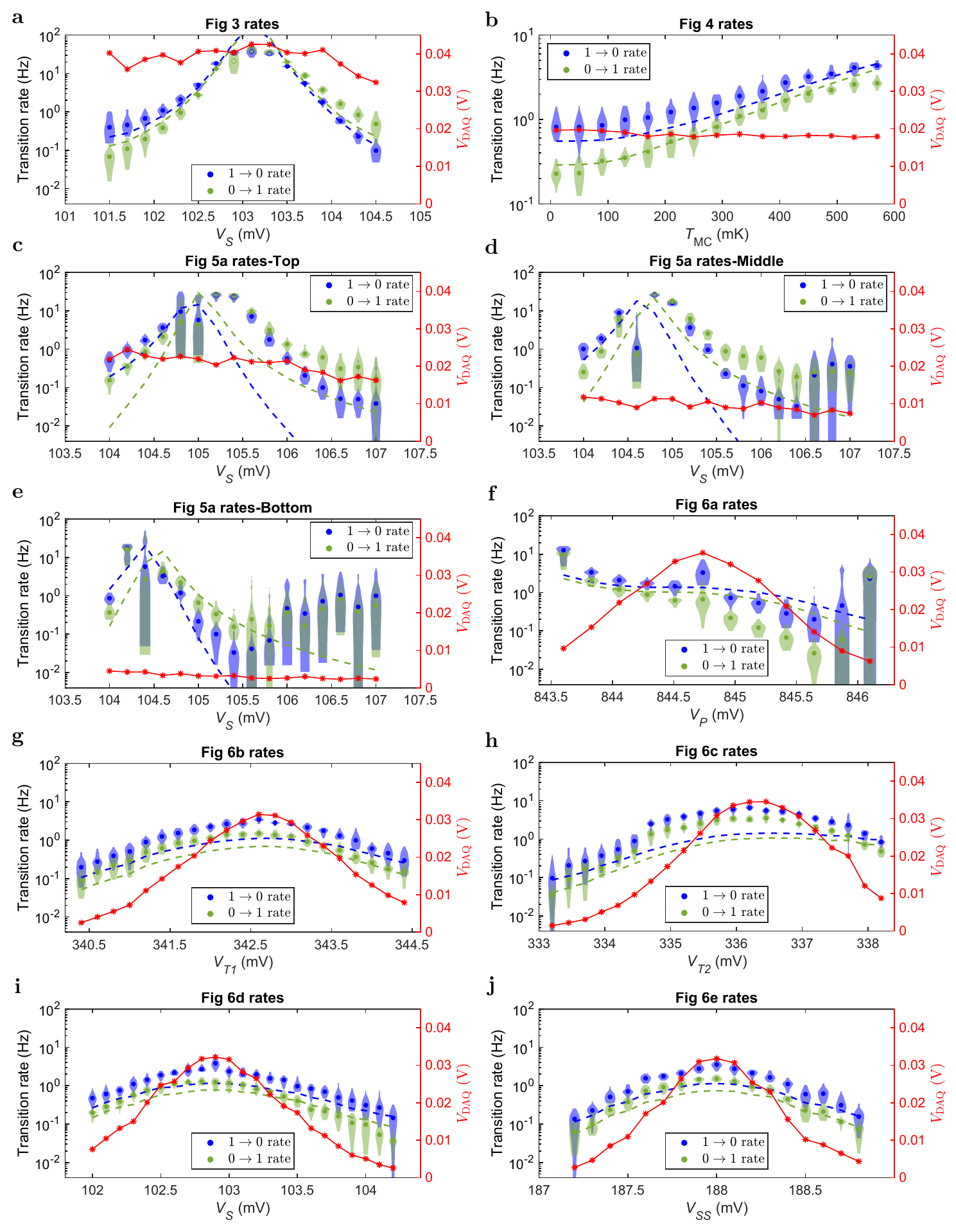}
  \caption{\textbf{Fit to measured TLF transition rates.} The tunneling plus thermal activation model shows qualitative agreement when fit to the full dataset. In blue (green) are the FHMM model-extracted transition rates $\Gamma_{01}$ ($\Gamma_{10}$), with the model fit given by the corresponding dashed curves. The red curve denotes the averaged DAQ signal, assumed to be proportional to the sensor-dot conductance. Open circles indicate data points excluded from the fit due to measurement bandwidth-induced aliasing.}
  \label{fig:FullFit}
\end{figure*}

Based on the data presented in the previous sections, a plausible model for this TLF is a vertically oriented double-well potential defect undergoing tunneling mediated not by electron-phonon coupling but by some other mechanism, perhaps involving coupling to electrons in the reservoirs. A horizontally oriented TLF near the edge of the plunger gate might also respond in a similar way to the gate voltages, although we might expect the TLF to respond to at least one other gate in this case. The strong local heating effect and the large amplitude of this TLF are also consistent with a defect localized near the dot under the plunger gate.

To understand the detailed dynamics of the TLF as a function of gate voltage variation and temperature variation, we develop a phenomenological model to capture the behavior of the TLF forward and reverse transition rates estimated from the experimental scans described in the previous sections.  We extract the rates from each experimental scan using a FHMM fit.  Each scan contains 30 repetitions, providing a distribution of forward and reverse rate estimates which we use to define a confidence interval for fitting the model.

In our model, we describe the TLF by the Hamiltonian $H_{\mathrm{TLF}} = (\epsilon \, \sigma_z + t_c \, \sigma_x)/2$, where $t_c$ denotes the tunnel coupling, $\epsilon$ the energy bias, and $\sigma_i$ are Pauli matrices. The voltages applied to gate electrodes are represented by the vector $\mathbf{V} = (V_{P},V_{S},V_{SS},V_{T1},V_{T2})$. We model the influence of these voltages on the TLF as acting on the detuning bias through
\begin{equation}
    \epsilon = \overrightarrow{\lambda} \cdot (\mathbf{V} - \mathbf{V}_{0}),
    \label{eq:LeverArm}
\end{equation}
where $\overrightarrow{\lambda}$ is a ``lever arm'' vector and $\mathbf{V}_{0}$ is some reference voltage in the hyperplane corresponding to the zero detuning bias point of the TLF.
Our model for the transition rate between state $j$ and $i$ is given by $\Gamma_{ij}^{\rm tot} = \Gamma_{ij}^{\rm tunneling} + \Gamma_{ij}^{\rm activated}$, where for an energy difference $\Delta E_{ij} \equiv E_{i} - E_{j}$,
\begin{multline}
    \Gamma_{ij}^{\mathrm{tunneling}} = \vert \langle E_{i} \vert \sigma_{z} \vert E_{j} \rangle \vert^{2} \left[ (1+n(\Delta E_{ji})) J\left( \Delta E_{ji} \right) \right.\\
    + \left.n(\Delta E_{ij}) J\left(\Delta E_{ij}\right)\right]
    \label{eq:Tunneling_rate}
\end{multline}
is the Fermi's Golden Rule \cite{phillips1987two} rate of tunneling from eigenstate $\vert E_{j} \rangle$ to $\vert E_{i} \rangle$ of the TLF, $J(x)$ is proportional to the density of bath states at the energy difference $E_{j}-E_{i}$, and $n(\Delta E) = (e^{\Delta E/k_{B}T}-1)^{-1}$ is the Bose-Einstein distribution with $k_{B}$ the Boltzmann constant and $T$ the temperature. 
The expression in Eq.~\eqref{eq:Tunneling_rate} is consistent with both a spin-boson model and a generalization of a model for electron-assisted tunneling \cite{phillips1987two}. 
We assume a power law dependence for the bath density of states $J(x)=c x^{b}$ with $c$ a free fit parameter and $b=-1$ fixed and chosen to agree with the voltage dependence of Fig.~\ref{fig:comp}c. We note that this dependence is qualitatively more consistent with electron-assisted tunneling but inconsistent with phonon-assisted tunneling. The coupling of the TLF to its environment is assumed to be mediated through the position dipole $\sigma_{z}$, such that $\langle E_{0} \vert \sigma_{z} \vert E_{1} \rangle = t_{c}/\sqrt{\epsilon^{2} + t_{c}^{2}}$. 
The detailed balance condition enforces $\Gamma_{ji}^{\mathrm{tunneling}} = \Gamma_{ij}^{\mathrm{tunneling}} e^{-\Delta E_{ji}/k_{B} T}$.
We allow for the presence of an additional thermally activated rate taking the Arrhenius form
\begin{equation}
    \Gamma_{ij}^{\mathrm{activated}} = \gamma_{0} e^{-(E_{b}-\Delta E_{ji}/2)/k_{B}T},
\end{equation}
where $E_{b}$ is a barrier energy and $\gamma_{0}$ is an ``attempt rate'' corresponding to the effective transition rate in the limit of vanishing barrier energy and energy difference $\Delta E_{ij}$. Note that this rate also obeys detailed balance. Hence, $\Gamma_{ji}^{\mathrm{tot}}/\Gamma_{ij}^{\mathrm{tot}} = e^{-\Delta E_{ji}/k_{B} T}$ and an asymptotically thermal distribution of TLF states is ensured. Our rationale for incorporating the sum of these two processes is to facilitate identifying whether environment-assisted tunneling or thermal activation over an energy barrier is the dominant driver of TLF dynamics \cite{Pourkabirian2014}. Finally, we model the temperature dependence of the TLF as follows:
\begin{equation}
    T_{\mathrm{TLF}} = \left( T_{\mathrm{MC}}^{1+\beta} + \kappa V_{\mathrm{DAQ}}\right)^{1/(1+\beta)},
\end{equation}
where $\kappa$ denotes a charge sensor heating lever arm and $\beta$ is the power law dependence on temperature of thermal conductivity in the sample.  The measurement signal $V_{\mathrm{DAQ}} = \mathcal{S}_{\mathrm{max}} - \mathcal{S}$ is assumed to be proportional to the sensor-dot conductivity, consistent with a circuit model for the rf reflectometry measurement and other similar measurement implementations \cite{Reilly2007FastRF}. This temperature dependence for heating follows from solving the heat equation under the assumption that the thermal conductivity obeys a power law $k(T) \propto T^{\beta}$ at low temperatures (for more details see the Supplemental Material \cite{SMnote}). We fix $\beta=3$ as the thermal conductivity exponent in all fits, consistent with what one may expect for bulk Si \cite{Thompson1961}. However, it is known that alloy disorder and interfaces can modify thermal conductivity significantly \cite{Toxen1961,Wang2010}. We find that $\beta$ is not tightly constrained by the fits here, but it is possible that a more detailed characterization of this heating effect may provide a better estimate of $\beta$ in this alloyed heterostructure and shed more light on heat transport through Si/SiGe devices.

\subsection{Fit results}
The model parameters we adjust for this fit include the heating lever arm $\kappa$ for each dataset, bare attempt rate $\gamma_0$, barrier energy $E_b$, and density of states prefactor $c$. In addition, we allow for a global bias offset to account for drift and few to tens of mV differences in device tuning voltages between measurements. We perform a minimization of the cost function $f(\lbrace \kappa_{\alpha} \rbrace, \gamma_0, E_b, c, \lbrace \Delta \epsilon_{\alpha} \rbrace) = \sum_{i} (\overline{r}_{i}^{\mathrm{data}} - r_{i}^{\mathrm{model}})^{2}/(\sigma_{i}^{\mathrm{data}})^{2}$
with respect to the aggregate set of measurements shown in Figs.~\ref{fig:comp}, \ref{fig:MCtemp}, \ref{fig:localheating}a, and \ref{fig:gate}(a-e), where $\overline{r}_{i}^{\mathrm{data}}$ ($\sigma_{i}^{\mathrm{data}}$) is the mean (standard deviation) of the transition rates $\Gamma_{10}$ and $\Gamma_{01}$, with $i$ an index for each experimental parameter setting. The comparison between the fits and the experimental data is shown in Fig. \ref{fig:FullFit}. The fit parameters we find are $(\kappa_{\mathrm{Fig3}},\kappa_{\mathrm{Fig4}},\kappa_{\mathrm{Fig5}},\kappa_{\mathrm{Fig6}}) = (0.2,0.02,5\times10^{-5},1.1) \ \mathrm{K^{4}/V}$, $\gamma_{0} = 81 \ \mathrm{Hz}$, $E_{b}=170 \ \mathrm{\mu eV}$, and the density of states factor is $J(\Delta E) \approx 30 (\Delta E)^{-1} \ \mathrm{Hz}$. To better control the optimization we fixed the lever arms to be $(\lambda_{P},\lambda_{S},\lambda_{SS},\lambda_{T1},\lambda_{T2}) = (-5.4,1.2,0,-0.8,1.5) \ \mathrm{\mu eV/mV}$ and the tunnel coupling to be $t_{c} = 2 \ \mathrm{\mu eV}$. These values are consistent with the results of separate detailed balance fits (see Supplemental Material \cite{SMnote}). In particular, these fits give $\lambda_{P}=-5.6_{-0.4}^{+0.6} \ \mathrm{\mu eV/mV}$, $\lambda_{T2}=1.3_{-0.1}^{+0.3} \ \mathrm{\mu eV/mV}$, and $t_{c} = 1.3_{-1.3}^{+1.6} \ \mathrm{\mu eV}$. The asymmetric confidence intervals represent the 95\% bounds of the $\chi^2$ (see Supplemental Material \cite{SMnote}). We note that our estimated thermal activation attempt rate $\gamma_{0}$ and activation barrier energy $E_{b}$ are of a similar order of magnitude to those measured in \cite{Kenyon2001}, where single-electron transistor heating and bias dependence were also observed.

\subsection{Interpretation}
When including both the tunneling and thermal activation rates $\Gamma_{ji}^{\mathrm{tunneling}}$ and $\Gamma_{ij}^{\mathrm{activated}}$, the model captures the qualitative voltage dependence and heating effects. However, the relatively poor goodness of fit to the full dataset suggests that either unmodeled parameter variation may need to be incorporated or a more sophisticated model developed. For example, we find that the best-fit heating lever arms $\kappa_{\alpha}$ vary significantly across experiments, though we would expect the conductance-dependent heating rates based on a simple Joule heating model to be relatively consistent across experiments. Proposals for more sophisticated TLF to single-electron transistor coupling models have previously been presented \cite{Kenyon2001} and warrant consideration in future work. 
Furthermore, we have explored incorporating temperature-dependent deviations from a simple Arrhenius rate \cite{Benderskii1993,Kohout2021} for thermal activation that significantly improve the model fit, though further improvements would be necessary to have a high confidence in model validity. 
Nonetheless, the qualitative agreement with our model provides plausible evidence that the TLF may be a bistable charge dipole in close proximity to the P gate electrode, heated by current through the sensor dot, and experiencing state transitions driven not by direct electron-phonon coupling but through some other mechanism such as coupling to electrons passing through the sensor dot.

\section{Conclusion}
%Through simple electrical transport measurements, we have characterized the response of individual TLFs to gate voltages and temperature variations in Si/SiGe quantum dots.
Through simple electrical transport measurements on quantum dots and analyses based on the Allan variance and factorial hidden Markov modeling, we have measured the properties of individual TLFs in Si/SiGe quantum dots. We have observed the dependence of the TLF switching times on gate voltages, temperature, and sensor-dot conductance through a local heating effect. A model for the TLF transition rates based on environment-assisted tunneling that accounts for gate voltage and local heating agrees at least qualitatively with the data. The ratio of transition rates $\Gamma_{01}/\Gamma_{10}$ satisfies the detailed balance condition, suggesting that the fluctuator statistics are those of a discrete charge degree of freedom in contact with a thermal bath. By evaluating the ratio of rates as a function of gate voltage and temperature, we can show that the energy bias of the TLF is predominantly coupled to the P gate electrode with a lever arm of magnitude approximately 5 $\mu$eV/mV, coupled to the T2 gate with a lever arm of opposite sign of magnitude approximately 1 $\mu$eV/mV, and only weakly coupled to the other gates.
%The relatively rapid vanishing of the transition rates (increase in switching time) when moving away from the zero-bias point appears to be inconsistent with the opposite dependence one would expect for a simple two-level charge trap coupled to phonons and may point to some other mechanism of environment-assisted tunneling. 

%Our work highlights the utility of the Allan variance in analyzing noise, especially in the case of a single dominant TLF~\cite{Czerwinski:09,burnett2019decoherence,yang2023locating}. In the multi-TLF regime, we show that modeling the system as a factorial hidden Markov model can provide information about the characteristics of multiple fluctuators. 

The voltage sensitivity of the switching times we observe potentially explains a variety of effects that have been previously noted, including the variation of the charge noise power spectrum between different sides of the same transport peak~\cite{connors2019low}, as well as the variation in noise levels between transport peaks~\cite{zwerver_qubits_2021,paquelet2023reducing} in the same devices. The prominence of individual TLFs also corroborates the picture in which deviations from a $1/f$-like noise spectrum are a result of a small number of fluctuators~\cite{connors2019low,connors2022charge}. 
The gate voltage sensitivity observed here prompts a number of questions. First, do the gate voltages used to manipulate the spin qubits themselves affect the charge noise of those qubits? The spectral density of charge noise is typically assumed to be constant in experiments, but this assumption may be violated if the noise sources depend on gate voltages. 
Second, is it possible to coherently control the TLFs with electric fields, as has been demonstrated in superconducting qubits~\cite{lisenfeld2010measuring}?
 
The temperature sensitivity and local heating effects observed here highlight the importance and challenge of accurately measuring TLF temperatures in order to determine their thermalization mechanisms. On the one hand, future work toward understanding and mitigating the local heating effect may enable a more accurate measurement of gate voltage sensitivities and the possibility of triangulating TLF locations. On the other hand, the local heating effect may also prove a valuable tool in characterizing a TLF's location, especially if the origin of the local heating can be determined. The local heating effect also calls into question the accuracy of sensor-dot proxy measurements for charge noise experienced by qubits.  

Our microscopic characterization of individual TLFs in Si/SiGe quantum dots illustrates how detailed measurements of individual fluctuators can open new pathways for the study of noise in quantum systems. These results complement traditional ensemble measurement techniques. Together, these methods promise to help diagnose and mitigate the noise sources that affect semiconductor spin qubits.

\section{Data Availability}
The processed data that support the findings of this study are available at \href{https://doi.org/10.5281/zenodo.10568688}{https://doi.org/10.5281/zenodo.10568688}. The
raw data are available from the corresponding author
upon reasonable request.

\section{Acknowledgments}
%We thank Lisa F. Edge of HRL Laboratories, LLC. for the epitaxial growth of the SiGe material.
We thank Lisa F. Edge of HRL Laboratories for growing the heterostructures and Elliot J. Connors for fabricating the devices used in this work.
This work was sponsored by the Army Research Office through Grant No. W911NF-17-1-0260 and W911NF-23-1-0115 and the Air Force Office of Scientific Research through Grant No. FA9550-23-1-0710. The views and conclusions contained in this document are those of the authors and should not be interpreted as representing the official policies, either expressed or implied, of the Army Research Office or the U.S. Government. The U.S. Government is authorized to reproduce and distribute reprints for Government purposes notwithstanding any copyright notation herein.

Sandia National Laboratories is a multimission laboratory managed and operated by National Technology and Engineering Solutions of Sandia, LLC, a wholly owned subsidiary of Honeywell International Inc., for the U.S. Department of Energy’s National Nuclear Security Administration under contract DE-NA0003525.

\section{Author Contributions}
F.Y., A.E., and J.M.N. formulated the experiment and carried out the measurements; D.A. analyzed the noise timeseries and performed the FHMM fits; R.V. implemented improvements to the NoMoPy code that facilitated the FHMM analysis; N.T.J. implemented the TLF model and performed supporting calculations; J.M.N. supervised the research.

%\bibliography{noisebib}


\end{document}

\title{Supplemental Material for \\ ``Characterization of individual charge fluctuators in Si/SiGe quantum dots"}

\author{Feiyang Ye}
\thanks{These authors contributed equally.}

\author{Ammar Ellaboudy}
\thanks{These authors contributed equally.}

\affiliation{Department of Physics and Astronomy, University of Rochester, Rochester, NY, 14627 USA}

\author{Dylan Albrecht}

\author{Rohith Vudatha}

\author{N. Tobias Jacobson}
\affiliation{Sandia National Laboratories, Albuquerque, NM, 87185 USA}

\author{John M. Nichol}
\email{john.nichol@rochester.edu}
\affiliation{Department of Physics and Astronomy, University of Rochester, Rochester, NY, 14627 USA}

%\date{\today}\\

\maketitle
	
%\date{\today}
	
%\section{Device}
%Device description

\tableofcontents

\section{Device fabrication}
Device 1 (Device 2) is fabricated on an undoped Si/SiGe heterostructure with a $2~\si{nm}$ ($4~\si{nm}$) Si capping layer. A 15-nm-thick $\mathrm{Al}_2\mathrm{O}_3$ gate dielectric is deposited via atomic layer deposition on both devices. Followed by the gate-oxide deposition, four (three) layers of overlapping aluminum gates are defined on Device 1 (Device 2) through a combination of photolithography and electron beam lithography steps. Aluminum gates are thermally evaporated on both devices. Each gate layer has a 3-nm-thick Al buffer layer deposited at a rate less than $0.1~\si{\AA/s}$, followed by the rest of aluminum deposited at a rate of about $2~\si{\AA/s}$. Between neighboring overlapping layers, a thin oxide layer is thermally formed to prevent gate-to-gate leakage.

\section{Lever arm and electron temperature measurements}
\begin{figure}[h]
\centering
\includegraphics[width=0.5 \textwidth]{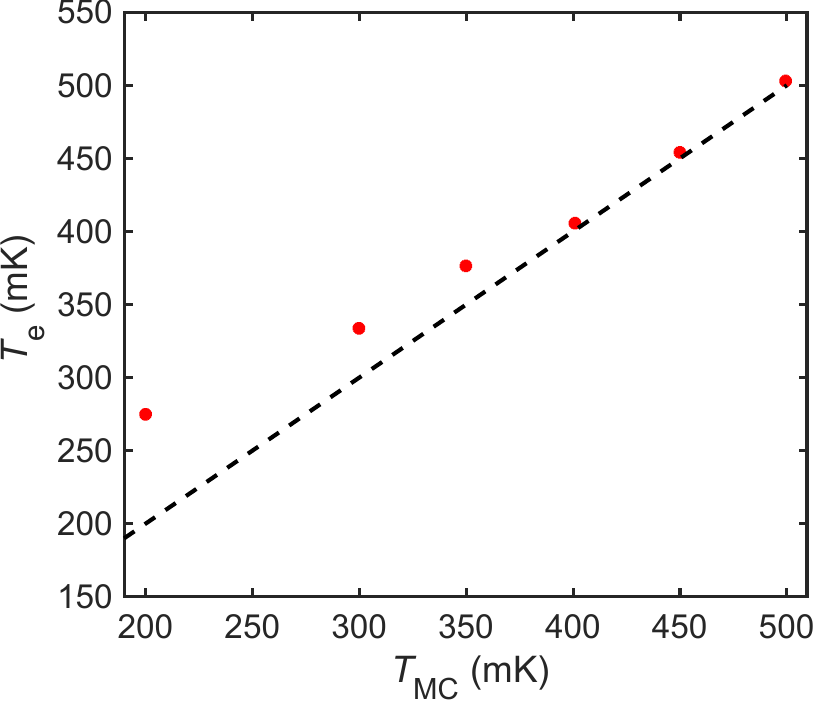}
\caption{
Electron temperature as a function of mixing chamber temperature (red) by fitting the transport peak shape of the right sensor dot in Device 1. The dashed line assumes $T_e = T_\mathrm{MC}$.
}
\label{fig:etemp}
\end{figure}

To find the lever arm $\alpha$, we raise the mixing chamber temperature $T_\mathrm{MC}$ to 500 mK and fit the transport peak shape to the predicted Coulomb blockade lineshape \cite{PhysRevB.44.1646}
\begin{equation}
    \mathcal{S} = C_1 \cosh^{-2} \qty(\frac{\alpha (V_P - V_0)}{2 k_B T_e}) + C_2,
\label{eq:CBpeak}    
\end{equation}
assuming pure thermal broadening. Here, $V_P$ is the sensor plunger gate voltage, $V_0$ is the plunger gate voltage at the center of the transport peak, $T_e$ is the electron temperature, and $C_i~(i=1,2)$ are fitting constants. Assuming $T_e=T_\mathrm{MC}=500~\si{mK}$, we extract lever arm $\alpha$ from a fit of the measured transport peak to Eq.~\eqref{eq:CBpeak}.

To find the relation between electron temperature and mixing chamber temperature, we measure a transport peak shape at each mixing chamber temperature, and fit the peak shape to get electron temperature. Figure~\ref{fig:etemp} plots how electron temperature changes vs. mixing chamber temperature. At elevated temperatures $T_\mathrm{MC} \ge 400~\si{mK}$, electron temperatures are approximately equal to mixing chamber temperatures. We did not adjust the tunnel coupling to minimize lifetime broadening so that we could measure the lever arm in the same tuning as the TLF measurements. Thus, the reported electron temperatures at low mixing chamber temperatures are likely to be significant overestimates.

\section{Other charge fluctuators}
\begin{figure}[h]
\centering
{\includegraphics[width=1 \textwidth,trim=5 5 5 5,clip]{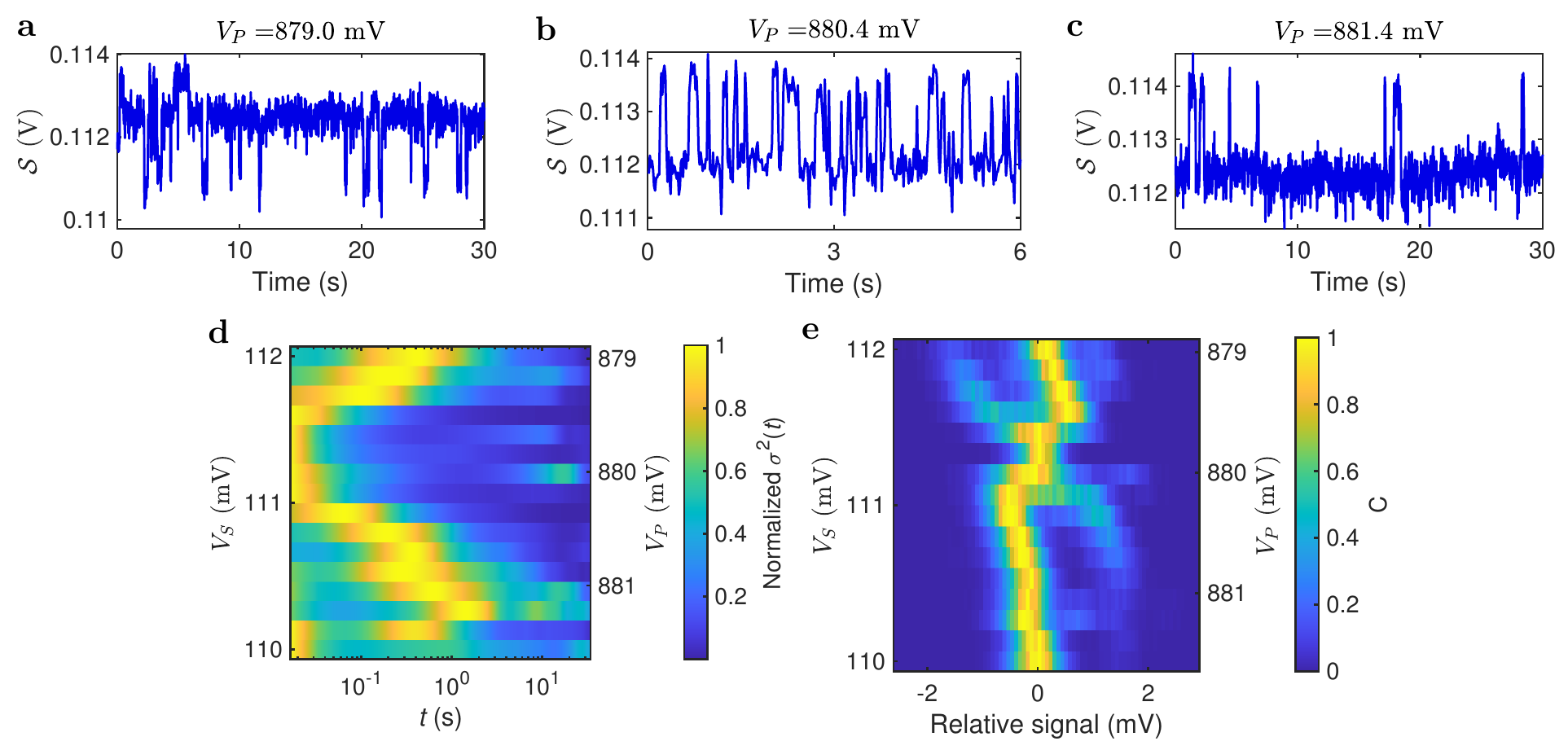}}
\caption{
\textbf{Voltage dependence of a TLF on the 3rd transport peak in Fig.~1b.}
\textbf{a} Time-trace of the TLF when the high signal state is the ground state.
\textbf{b} Time-trace of the TLF when switching time is close to its minimum.
\textbf{c} Time-trace of the TLF when the low signal state is the ground state.
\textbf{d} Switching times vs. screening and plunger gate voltages. The switching time goes to a minimum when $V_P \approx 880~\si{mV}$.
\textbf{e} Occupation of the TLF states at different gate voltages. The most likely state changes on either side of $V_P \approx 880~\si{mV}$.
}
\label{figS:volt}
\end{figure}

\begin{figure}[h]
\centering
{\includegraphics[width=0.9 \textwidth,trim=0 0 0 0,clip]{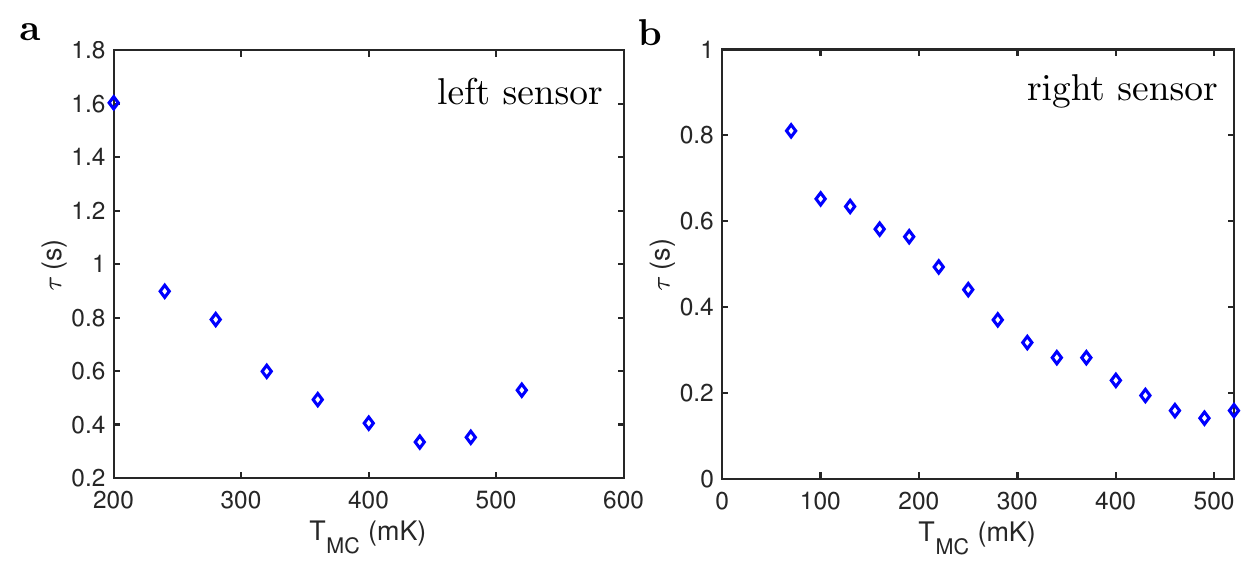}}
\caption{
\textbf{Temperature dependence.}
\textbf{a} Temperature dependence of a TLF measured in Device 1 with the left charge sensor. 
We sweep $T_\text{MC}$ from $200~\si{mK}$ to $520~\si{mK}$ with a $40~\si{mK}$ increment.
\textbf{b} Temperature dependence of the TLF measured in Device 1 on the 3rd transport peak of the right charge sensor. We sweep $T_\text{MC}$ from $70~\si{mK}$ to $550~\si{mK}$ with a step of $30~\si{mK}$.
}
\label{figS:T}
\end{figure}

\begin{figure}[h]
\centering
{\includegraphics[width=1 \textwidth,trim=0 0 0 0,clip]{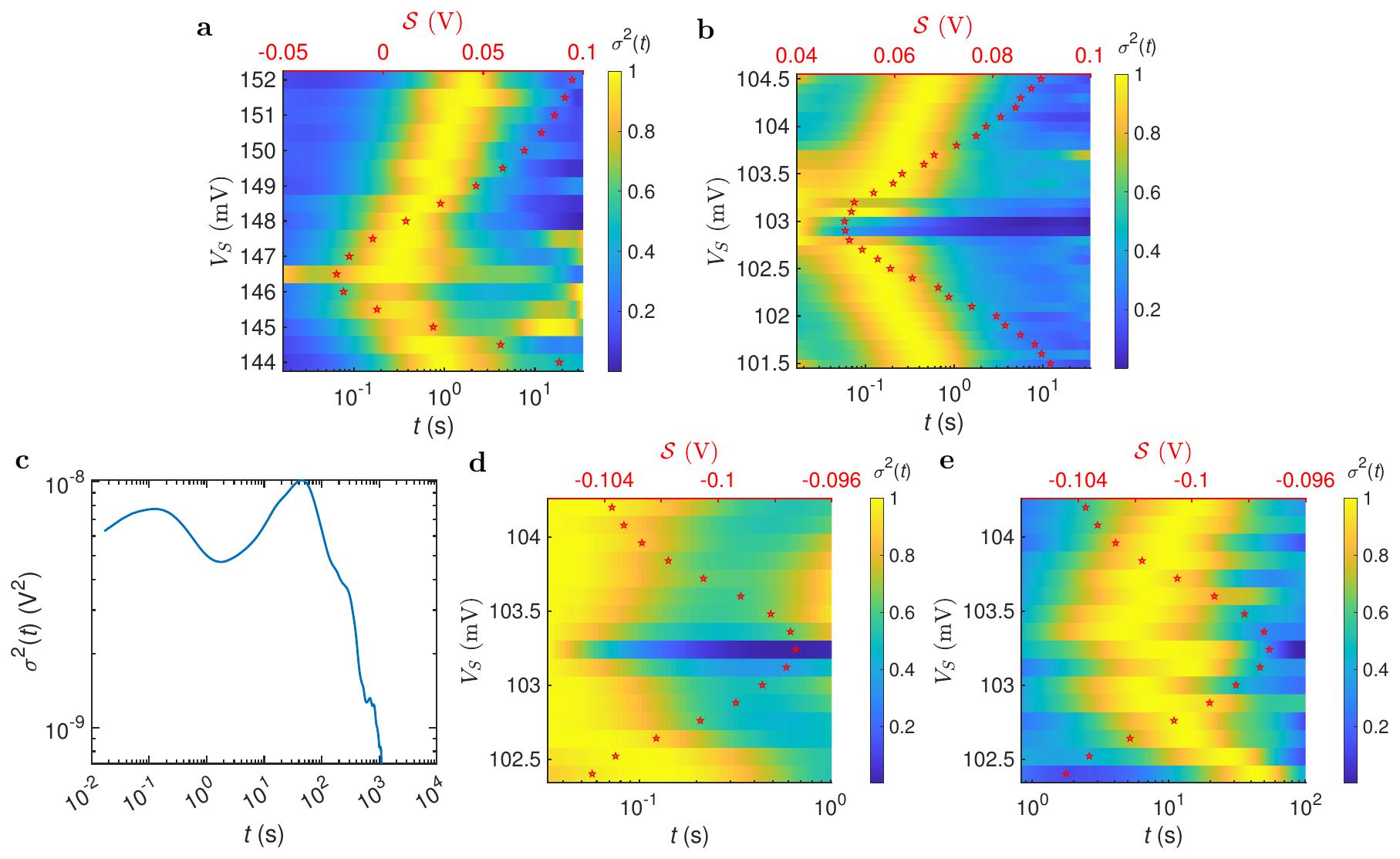}}
\caption{
\textbf{Local heating.}
\textbf{a} Local heating measured in Device 2.
\textbf{b} Local heating effect measured in Device 1 on the 3rd transport peak of the right charge sensor.
\textbf{c}-\textbf{e} Local heating observed in two TLFs measured in Device 1 on the left charge sensor.
\textbf{c} An example of the Allan variance measured in Device 1 on the left sensor dot. The two peaks of the Allan variance suggest the presence of two TLFs, one with a short switching time and the other with a long switching time. 
\textbf{d} Local heating in the fast TLF. We select the switching time window below $10^0~\si{s}$ to focus on the fast TLF. 
\textbf{e} Local heating for the slow TLF. We select the switching time window between $10^0$ and $10^2~\si{s}$ to focus on the slow TLF. In \textbf{d} and \textbf{e}, Allan variances $\sigma^2(t)$ are normalized by their maximum values within the selected time windows. The local heating affects both TLFs simultaneously.
}
\label{figS:LH}
\end{figure}

\begin{figure}[h]
\centering
{\includegraphics[width=1 \textwidth,trim=5 5 5 5,clip]{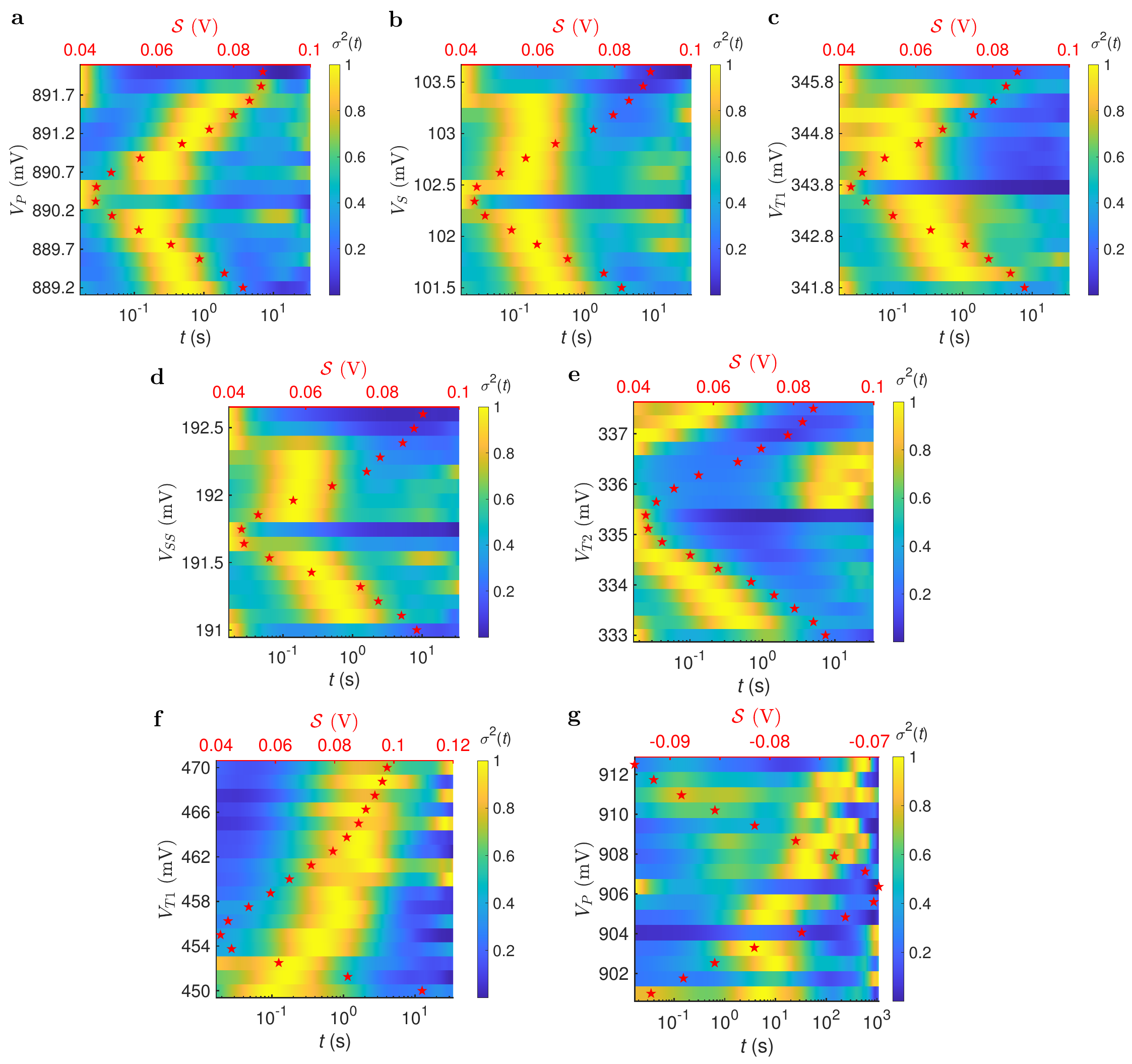}}
\caption{
\textbf{Gate-voltage sensitivity.}
\textbf{a}-\textbf{e} Switching times and mean sensor signals vs. different sensor gate voltages for a different TLF measured in Device 1 on the 3rd transport peak of the right sensor. At the top and bottom of the transport peak, the sensor loses its sensitivity to charge noise. 
The TLF switching time responds nearly symmetrically to most gates, except the sensor screening gate SS and the tunneling gate T2.
In \textbf{e}, the strong asymmetry about the center of the transport peak suggests that the tunneling gate T2 couples most strongly to the TLF. Near $\text{T2}=336~\si{mV}$, the switching time of the TLF in focus here decreases rapidly such that the TLF cannot be detected, and other fluctuators with switching times on the order of $10~\si{s}$ mainly contribute to the sensor signals.
\textbf{f} Tunneling gate T1 voltage dependence of a charge fluctuator measured in Device 2. The monotonic change of the switching time about the center of the transport peak indicates that the detuning effect induced by the T1 gate is stronger than the local heating effect.
\textbf{g} Plunger gate voltage dependence of a fluctuator on the left sensor dot in Device 1. The monotonic trend of switching times across the conductance peak suggests that the plunger gate couples strongly to the fluctuator.
}
\label{figS:gate}
\end{figure}

\begin{figure}[h]
\centering
{\includegraphics[width=0.75 \textwidth,trim=0 0 0 0,clip]{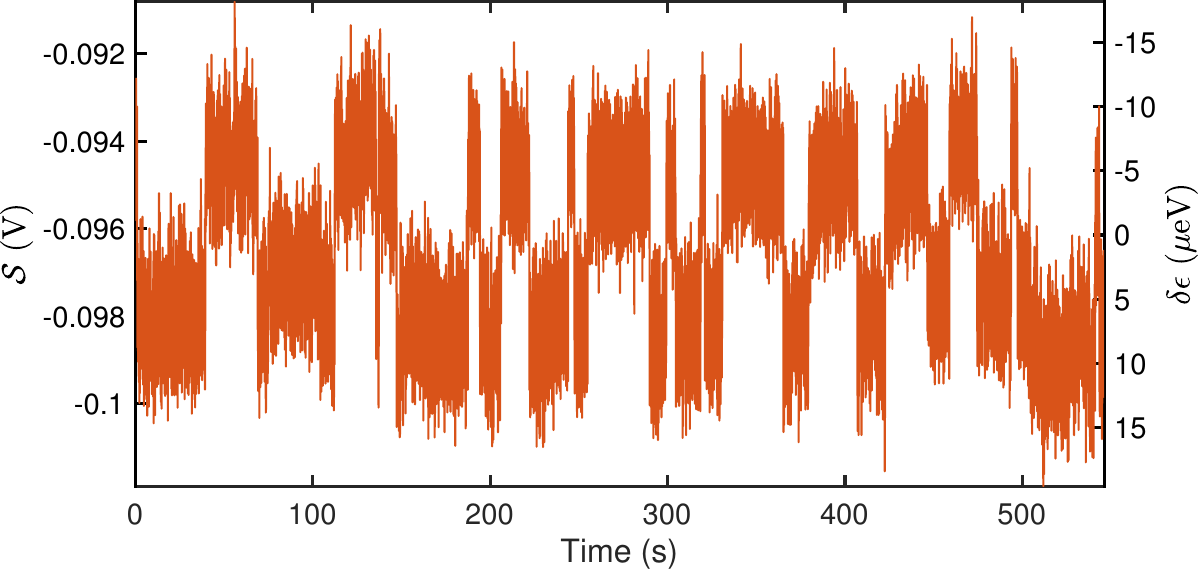}}
\caption{
\textbf{Large-amplitude TLF.}
We acquire a time series on the left sensor dot on Device 1, and observe a large amplitude TLF, similar to the one reported in the main text. 
}
\label{figS:Tseries}
\end{figure}

To confirm the repeatability of the trends discussed in the main text, we characterize other TLFs in Device 1 and Device 2.
We find that other TLFs show similar behaviors including voltage dependence (Fig.~\ref{figS:volt}), temperature dependence (Fig.~\ref{figS:T}), local heating (Fig.~\ref{figS:LH}), and gate-voltage sensitivity (Fig.~\ref{figS:gate}). In addition, we observe another TLF in Device 1 with an amplitude similar to the one discussed in the main text (Fig.~\ref{figS:Tseries}).

% voltage dependence of another TLF

\newpage

\newpage
% local heating effect

\newpage
% gate-voltage sensitivity

\newpage
% Large-amplitude TLF

\section{Bias analysis}
To extract the bias between the $0$ state and the $1$ state of the TLF defined in Eq.~(1) in the main text, we fit a histogram of reflectometry signals to a sum of two Gaussians $G(\mathcal{S}) = g_0(\mathcal{S}) + g_1(\mathcal{S})$, where
\begin{equation}
    g_i(\mathcal{S}) = \frac{A_i}{\sqrt{2\pi \sigma^2_i}} \exp\qty(- \frac{\qty(\mathcal{S}-\mu_i)^2}{2 \sigma^2_i}).
%+ \frac{A_R}{\sqrt{2\pi \sigma^2_R}} \exp\qty(- \frac{\qty(v-\mu_R)^2}{2 \sigma^2_R}),
\end{equation}
We assume that the reflectometry signals $\mathcal{S}$ of each charge state follow a Gaussian distribution $g_i(\mathcal{S})$ with amplitude $A_i$, mean $\mu_i$, and standard deviation $\sigma_i$, respectively, where $i=0,1$ indicates the charge state of the fluctuator \cite{connors2020rapid}.
The threshold signal $\mathcal{S}_t$ separating the two states is chosen to maximize the readout fidelity
defined as
\begin{equation}
    F = \frac{1}{2} \qty(\frac{\int_{-\infty}^{\mathcal{S}_t} g_0(\mathcal{S}) d\mathcal{S}}{\int_{-\infty}^{+\infty} g_0(\mathcal{S}) d\mathcal{S}} + \frac{\int_{\mathcal{S}_t}^{+\infty} g_1(\mathcal{S}) d\mathcal{S}}{\int_{-\infty}^{+\infty} g_1(\mathcal{S}) d\mathcal{S}}).
\end{equation}
After determining the threshold signal $\mathcal{S}_t$, we compute the ``bias" defined in Eq.~(1) using the relation
\begin{equation}
    \mathcal{B} \equiv \frac{N_0}{N_1} = \frac{\int_{-\infty}^{\mathcal{S}_t} g_0(\mathcal{S}) d\mathcal{S}}{\int_{\mathcal{S}_t}^{+\infty} g_1(\mathcal{S}) d\mathcal{S}},
\end{equation}
where each integral counts the population of an individual charge state $N_i~(i=0,1)$.

\section{Local heating at an elevated mixing chamber temperature} 
\begin{figure}[h]
\centering
{\includegraphics[width=1 \textwidth,trim=0 0 0 0,clip]{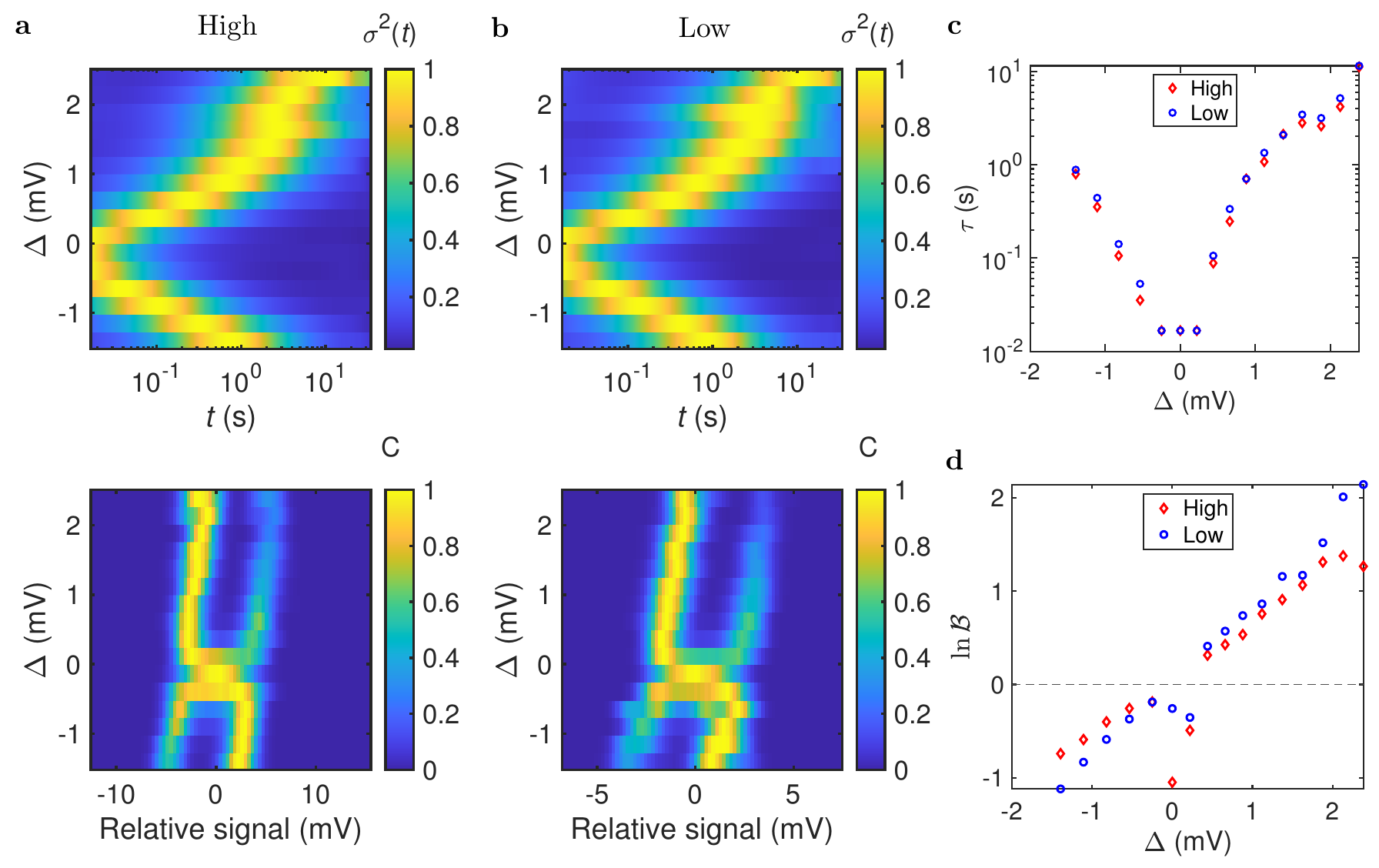}}
\caption{
\textbf{Local heating effect at an elevated temperature.}
We heat up the mixing chamber temperature to $200~\si{mK}$.
In \textbf{a} and \textbf{b}, the top (bottom) panel shows the switching time (histogram of the sensor signals) vs. gate voltage for \textbf{a} high power, and \textbf{b} low rf power.
\textbf{c} Switching time vs. gate voltage with high and low rf powers. The TLF switching times tend to decrease with increased current through the sensor dot.
\textbf{d} The natural logarithm of the bias vs. gate voltage with high and low rf powers, where the horizontal line indicates equal bias $\mathcal{B}=1$. Similar to the analysis of the main text, we calculate that the effective temperature ratio between high and low rf power is $1.35 \pm 0.17$, which is nearly half of the value $2.76 \pm 0.20$ at base temperature $T_\text{MC}=10~\si{mK}$.
}
\label{figS:200mK}
\end{figure}

Heating up the mixing chamber temperature to $200~\si{mK}$, we use the same gate voltages to hold the sensor dot conductance fixed while changing the current through the sensor dot with different rf powers (Fig.~\ref{figS:200mK}). Figures~\ref{figS:200mK}a,b show the gate-voltage dependent switching times and histograms of the sensor signals with high and low rf powers. Identical to what we see in Fig.~5, the TLF switching times tend to decrease with increased current through the sensor dot (Fig.~\ref{figS:200mK}c). Figure~\ref{figS:200mK}d plots the computed bias from the histograms as a function of gate voltage with high and low rf powers, from which we calculate that the effective temperature ratio between high and low rf power is $1.35 \pm 0.17$, which is nearly half of the value $2.76 \pm 0.20$ at base temperature $T_\text{MC}=10~\si{mK}$. When the mixing chamber becomes a major heating source at elevated temperatures, the local heating effect gets less pronounced.

% Local heating at 200 mK

 %%%% Start of modeling section
\section{Compensated voltage sweep trajectories}
\begin{figure*}[ht]
\centering
  \includegraphics[width=0.7 \textwidth,trim=0 0 0 0,clip]{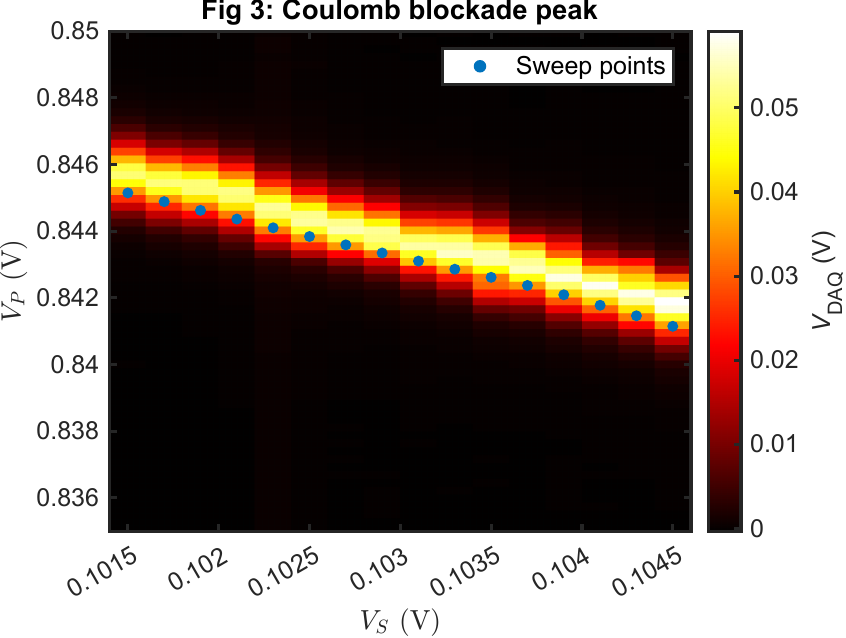}
  \caption{\textbf{Trajectory of voltage sweep for Fig 3.} Sweep points in the ($V_{P}$, $V_{S}$) plane are shown in blue, with the voltage-dependent charge sensor Coulomb blockade peak shown. This compensation of ($V_{P}$, $V_{S}$) ensures that the charge sensor signal is approximately constant throughout the voltage sweep.}
  \label{fig:Fig2_sweep_trajectory}
\end{figure*}

\begin{figure*}[ht]
\centering
  \includegraphics[width=0.7 \textwidth,trim=0 0 0 0,clip]{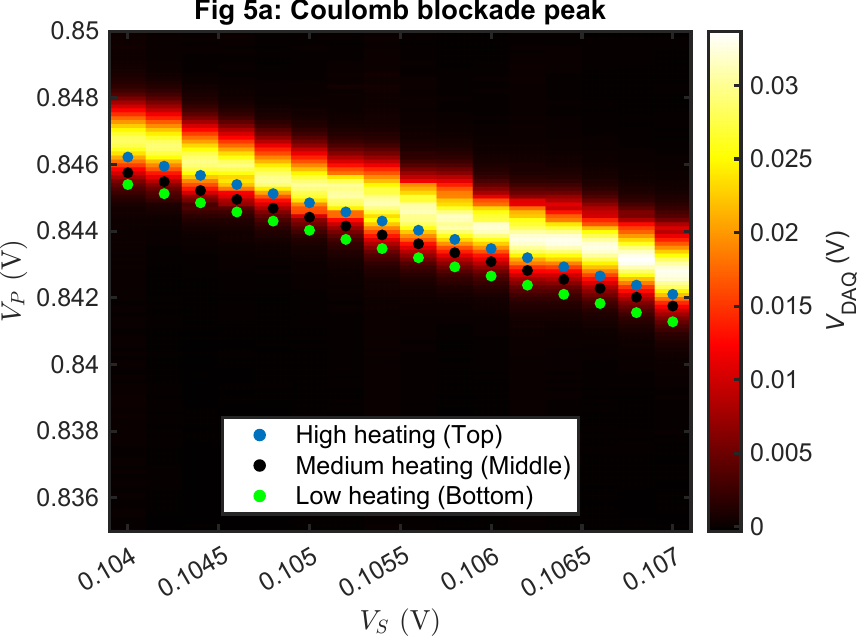}
  \caption{\textbf{Trajectory of voltage sweep for Fig 5a.} Sweep points in the ($V_{P}$, $V_{S}$) plane are shown in dots, with the voltage-dependent charge sensor Coulomb blockade peak shown. This compensation of ($V_{P}$, $V_{S}$) ensures that the charge sensor signal is approximately constant throughout the voltage sweep, with different trajectories sitting at different charge sensor signal levels corresponding to different CS heating levels.}
  \label{fig:Fig4_sweep_trajectory}
\end{figure*}

In Figs. \ref{fig:Fig2_sweep_trajectory} and \ref{fig:Fig4_sweep_trajectory}, we show the operating voltages for the voltage sweeps of Figs. 3 and 5a in relation to the SET Coulomb blockade peak. In both cases, we vary gate voltages $V_P$ and $V_S$ along a diagonal slope such that the charge sensor conductance is approximately constant.

\section{Model selection for three Coulomb blockade peaks}
\label{sec:model_selection}
In the main text, Fig.~2 illustrates how qualitatively different charge noise manifests in the vicinity of distinct Coulomb blockade peaks. Here, we discuss our method for inferring a multi-TLF model for the charge noise environment of at these three Coulomb blockade peaks through a model selection analysis based on maximum likelihood-based cross validation using \emph{NoMoPy}, a code for working with factorial hidden Markov models (FHMM) \cite{Albrecht2023}.

Each dataset of Fig. 2 contains 30 repetitions.  For the number of TLFs, $d$, ranging from 1-4, we fit a FHMM to each repetition while testing the performance on a held-out repetition. The purpose of this technique of cross-validation is to find the most appropriate number of fluctuators to model the data and to avoid overfitting -- in the event of overfitting one would expect the model performance to degrade when tested against the unseen data of the test set, while in contrast the performance would continue to increase on the training set.  
We use the log likelihood to quantify the performance of the fits \cite{Albrecht2023}. 
In practice, we tend to see a saturation of the log likelihood on the test set when adding more fluctuators to the model.  In addition, we note that when continuing to increase the number of fluctuators present in the model, the fitting will estimate fluctuators with tinier weights and more uncertain, higher transition frequencies, attempting to capture the noise in greater detail.

\begin{figure}[ht]
\centering
  \includegraphics[width=0.5 \textwidth,trim=0 0 0 0,clip]{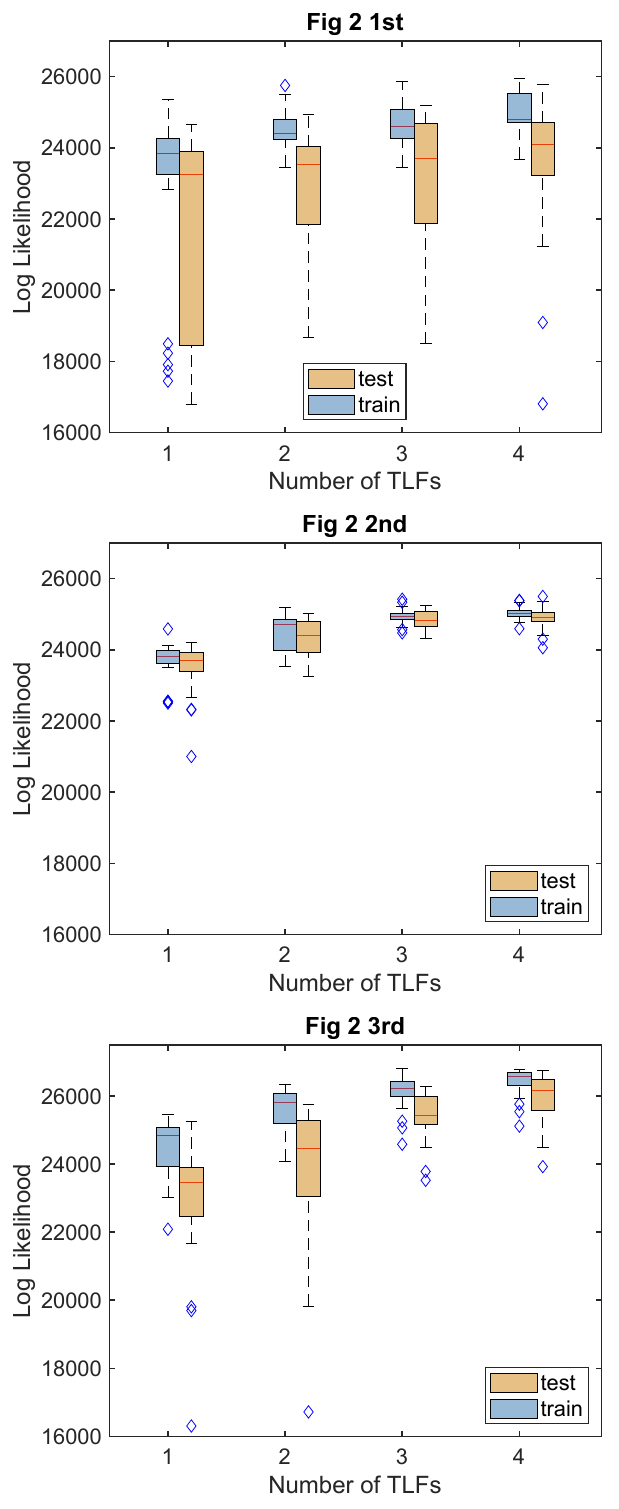}
  \caption{\textbf{Model selection for the three datasets of main text Fig.~2.} The first figure doesn't show much improvement going to higher numbers of TLFs. The second figure shows saturation of the test log-likelihood around three TLFs. The third figure appears to continue to improve possibly beyond four TLFs.}
  \label{fig:Model_selection}
\end{figure}

This procedure resulted in 30 trained model likelihoods and 30 test-set likelihoods, for each value of $d$. These likelihoods are plotted in Fig.~\ref{fig:Model_selection}.  By looking at the spread and trend of the score on the test likelihood, we selected the number of TLFs to include in the model for each dataset.  The scores for the 1st figure remain mostly flat, so we picked $d=1$.  For the 2nd plot, there is no improvement going from $d=3$ to $d=4$, so we chose a model with $d=3$. Finally, for the 3rd dataset, there appeared to be no saturation, so we chose the maximum tested, $d=4$.  All runs were completed on 32 cores of 2 nodes of Sandia National Laboratories' SkyBridge cluster, using the Python package \emph{dask} for parallelization, taking about 6 hrs in total.  This could easily be further parallelized to reduce the total computation time to $\sim$ 1~hr.

\section{Model for heating}
In this section, we describe a simple model for charge sensor heating and derive a dependence for TLF temperature on heating power that accounts for temperature-dependent thermal conductivity at cryogenic temperatures. In our model for the experiment, we assume that the effective temperature of the TLF's environment may change due to (1) varying mixing chamber temperature $T_{\mathrm{MC}}$ in the dilution refrigerator (e.g. lattice temperature) or (2) dissipating current through the charge sensor (CS) by changing its conductance $G_{\mathrm{CS}}$ or source-drain bias voltage. Below, we consider the effective temperature of the TLF due to either changing the mixing chamber temperature or modifying the CS conductance at fixed source-drain bias.

\subsection{Temperature-dependent thermal conductivity}
\begin{figure}[ht]
\centering
  \includegraphics[width=0.65 \textwidth,trim=0 0 0 0,clip]{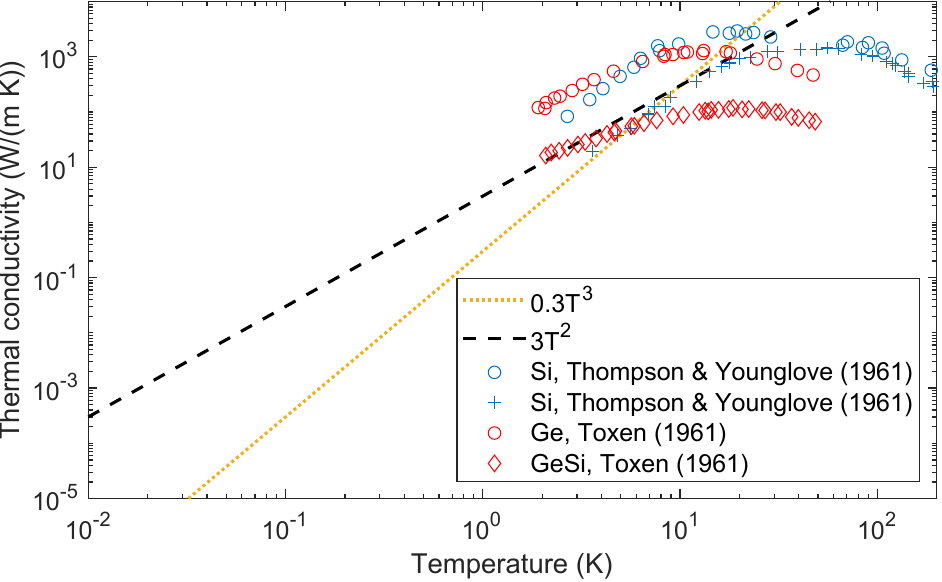}
  \caption{Thermal conductivity of Si, Ge, and a GeSi alloy. The red and blue points are taken from the measurements of \cite{Toxen1961,Thompson1961}. For reference, we also plot $T^{2}$ and $T^{3}$ dependencies. The orange dotted curve with $k \propto T^{3}$ is the thermal conductivity assumed for the example finite element simulations below.}
  \label{fig:Thermal_conductivity}
\end{figure}

It is well known that the thermal conductivity of Si, Ge, and SiGe alloys is strongly temperature-dependent at cryogenic temperatures \cite{Toxen1961,Thompson1961}. This temperature dependence has dramatic effects on the heating response of a device at cryogenic temperatures. Consider the steady-state heat equation given by
\begin{equation}
    \label{eq:HeatEqn}
    -\nabla \cdot (k(u) \nabla u) = \dot{q}(\mathbf{r}),
\end{equation}
where $u(\mathbf{r})$ is a spatially-varying temperature, $k(u)$ is a temperature-dependent thermal conductivity, and $\dot{q}(\mathbf{r})$ is a heating power density. An important consideration of systems at cryogenic temperatures is that the thermal conductivity $k$ is highly temperature-dependent, rapidly vanishing towards zero temperature. Shown in Fig. \ref{fig:Thermal_conductivity} is the thermal conductivity dependence for Si, Ge, and a GeSi alloy measured in \cite{Toxen1961,Thompson1961}. Alloy composition may strongly influence the thermal conductivity \cite{Wang2010,Cheaito2012}, so significant variability should be expected for thermal conductivity in heterostructures with alloy disorder and interfaces.

Given a thermal conductivity taking the power law form $k(T) = \alpha T^{\beta}$, we can solve the heat equation in the following way. Note that we can rewrite the heat equation Eq. (\ref{eq:HeatEqn}) as
\begin{equation}
    -\nabla \cdot (u^{\beta} \nabla u) = \dot{q}(\mathbf{r})/\alpha.
\end{equation}
Defining a new variable $f(u) = u^{\beta+1}/(\beta+1)$, we can see that $f$ obeys the linear heat equation
\begin{equation}
    -\nabla^{2} f = \dot{q}(\mathbf{r})/\alpha.
    \label{eq:Modified_heat_equation}
\end{equation}
Hence, all we need to do is solve the linear heat equation for $f(\mathbf{r})$ and then transform into the temperature variable via $u(f)=((\beta+1) f)^{1/(\beta+1)}$. Given the linearity of Eq. (\ref{eq:Modified_heat_equation}), we can account for Dirichlet boundary conditions (BCs) (assumed here to be set by the lattice or mixing chamber temperature) straightforwardly. The BCs for Eq. (\ref{eq:Modified_heat_equation}) are given by $f_{\mathrm{BC}} = T_{\mathrm{MC}}^{\beta + 1}/(\beta + 1)$, from which we obtain
\begin{equation}
    u[\frac{\lambda \dot{q}}{\alpha}, T_{\mathrm{MC}}](\mathbf{r}) = \left( T_{\mathrm{MC}}^{\beta+1} + \lambda (u[\frac{\dot{q}}{\alpha},0](\mathbf{r}))^{\beta + 1}\right)^{1/(\beta + 1)},
    \label{eq:Nonlinear_heat_eqn_solution}
\end{equation}
where $u[\dot{q}/\alpha,T_{\mathrm{MC}}](\mathbf{r})$ is the temperature at coordinate $\mathbf{r}$ for heating power density $\dot{q}(\mathbf{r})$, thermal conductivity $k(T) = \alpha T^{\beta}$, and mixing chamber temperature $T_{\mathrm{MC}}$. The dimensionless parameter $\lambda$ allows for an arbitrary global rescaling of the heating power density, such that given some solution $u[\dot{q}/\alpha,0](\mathbf{r})$ at a reference heating power density $\dot{q}$ we can immediately determine the solution at any other heating power and choice of temperature $T_{\mathrm{MC}}$. Note that in the limit of a temperature-independent thermal conductivity, $\beta \to 0$, we obtain the expected linear response of temperature on heating power.

Assuming that the heating power is proportional to the SET conductivity $G_{\mathrm{CS}}$ via Joule heating as $P_{\mathrm{SET}} = V_{\mathrm{SD}}^{2} G_{\mathrm{CS}}$, we finally obtain the following temperature dependence at the TLF location as a function of heating and mixing chamber temperature:

\begin{equation}
    T(T_{\mathrm{MC}},G_{\mathrm{CS}}) = \left( T_{\mathrm{MC}}^{\beta + 1} + \kappa_{G} G_{\mathrm{CS}} \right)^{1/(\beta + 1)}.
    \label{eq:T_dependence}
\end{equation}
We note that this functional form agrees well with the inferred TLF temperature variation due to SET self-heating measured in Ref. \cite{Gustafsson2013} if one were to take $\beta \approx 2.5$.
% In our preliminary fits shown below to the heating power of the charge sensor, we assume $\beta=2$ for the exponent.

\begin{figure}[ht]
\centering
  \includegraphics[width=0.6 \textwidth,trim=5 5 5 5,clip]{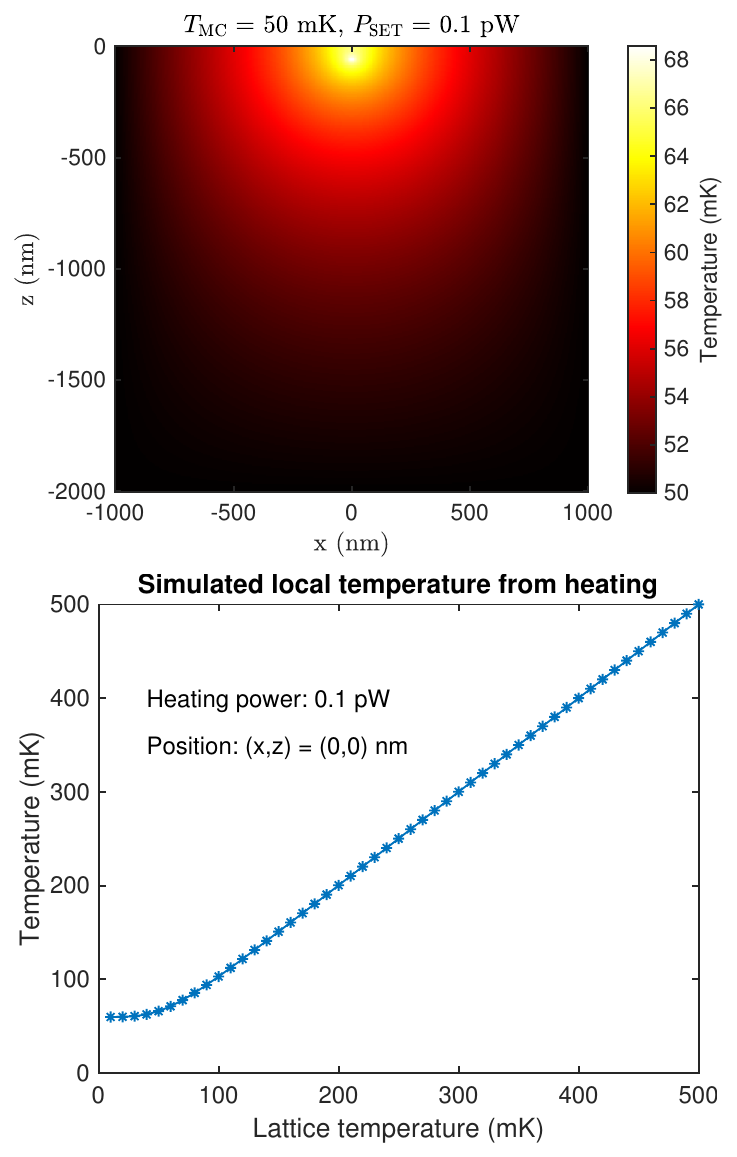}
  \caption{\textbf{Finite element simulations of charge sensor heating.} (top) Simulation in 2D of the temperature profile in the $x$-$z$ plane, assuming heating localized to 60 nm below the surface of the semiconductor and uniform along the $y$-axis. The plotted temperature is in units of mK. The assumed boundary conditions at the edge of the simulation domain are a Dirichlet condition $T_{\mathrm{MC}} = 50 \ \textrm{mK}$ on the bottom and sides and a zero-flux condition $\partial_{z} u = 0$ at the top to model no heat flow out of the top of the device. The total heating power is assumed to be 0.1 pW with a Gaussian heating power density of $\sigma_{x} = 10 \ \mathrm{nm}$ and $\sigma_{z} = 4 \ \mathrm{nm}$ centered at $(x,z)=(0,-60) \ \mathrm{nm}$ and assumed uniform along a channel length of 100 nm along the $y$-axis. In this simulation, we assume a thermal conductivity of $k(T) = 0.3 T^{3}$. (bottom) Variation of temperature at the top of the simulation domain as $T_{\mathrm{MC}}$ is varied.}
  \label{fig:FEM_simulation}
\end{figure}

\begin{figure}[ht]
\centering
  \includegraphics[width=0.7 \textwidth,trim=0 0 0 0,clip]{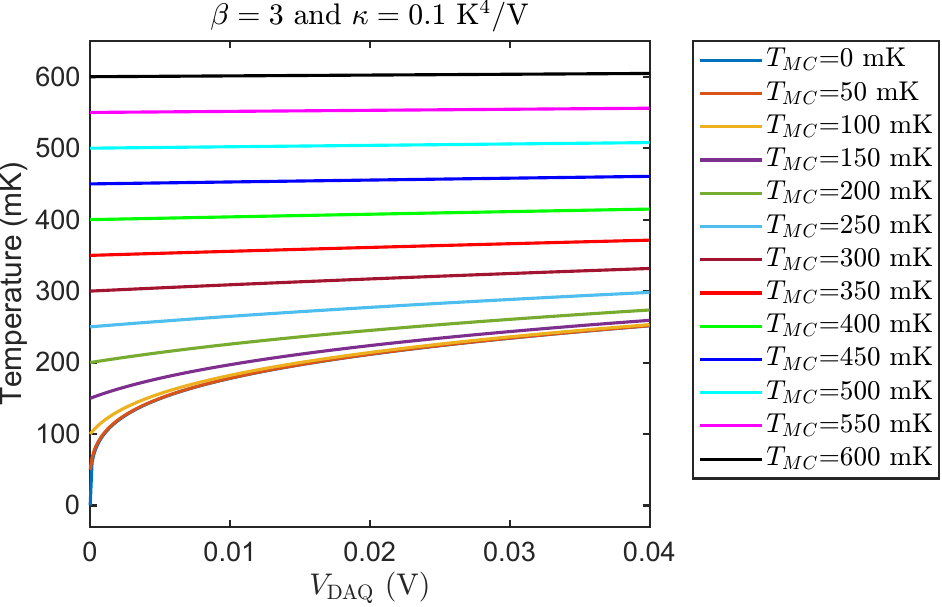}
  \caption{\textbf{Charge sensor heating effect.} Estimated temperature of the TLF as a function of DAQ signal, a proxy for conductance through the SET $G_{\mathrm{CS}}$, for various values of the mixing chamber temperature $T_{\mathrm{MC}}$. The functional form is $T_{\mathrm{TLF}} = (T_{\mathrm{MC}}^{1+\beta} + \kappa V_{\mathrm{DAQ}})^{1/(1+\beta)}$, where here we assume a heating lever arm value of $\kappa = 0.1 \ \mathrm{K^{4}/V}$ with $\beta = 3$. The dynamic range of DAQ signals shown here is consistent with the measurements.}
  \label{fig:CS_heating_effect}
\end{figure}

In Fig.~\ref{fig:FEM_simulation}, we show an example temperature profile arising from localized heating. The thermal conductivity we assume for this simulation is $k(T) = 0.3 \ T^{3} \ \left[\mathrm{W/(m \ K)}\right]$, corresponding to the dotted orange line of Fig.~\ref{fig:Thermal_conductivity}. This simulation in 2D assumes a total heating power of 100 fW distributed over a 100 nm long channel along the $y$-axis and localized in the $x$-$z$ plane. Note that a more realistic description would account for the three-dimensional nature of the problem, local current density corresponding to the SET geometry, spatially-varying thermal conductivity due to the heterostructure, and boundary effects due to the gate dielectric and gate electrodes on the top of the sample. However, this simple simulation and Fig.~\ref{fig:CS_heating_effect} illustrate the relevant physics arising from a temperature-dependent thermal conductivity.

In Fig.~\ref{fig:CS_heating_effect}, we show the estimated heating effect at the TLF location due to the variation in the charge sensor conductance $G_{\mathrm{CS}}$ as reflected in the DAQ signal $V_{\mathrm{DAQ}}$. Note that the range of DAQ signal shown here is representative of the dynamic range observed in the experiment. In this plot, we can see that the relative influence of the heating effect is stronger at lower mixing chamber temperatures, as should be expected due to the suppressed thermal conductivity at lower temperatures that establishes a stronger thermal gradient for the same total power dissipation.

\section{Model for TLF transition rates}
As described in the main text, our model for the TLF is as a two-level system having the Hamiltonian $H_{\mathrm{TLF}} = (\epsilon \sigma_{z} + t_{c} \sigma_{x})/2$ in the basis of localized states $\lbrace \vert 0 \rangle, \vert 1 \rangle \rbrace$, where $t_{c}$ denotes a tunnel coupling and $\epsilon$ an energy bias. This TLF is part of a larger system, the SiGe/Si/SiGe device, consisting of a semiconductor heterostructure, metal gates, gate dielectrics, and nearby two-dimensional electron gas (2DEG) in the Si quantum well. To account for the interactions between this TLF and its environment, we consider the TLF to be in contact with a bath at some temperature $T$. The voltages applied to the gate electrodes controlled in this experiment are represented by the vector $\mathbf{V} = (V_{P},V_{S},V_{SS},V_{T1},V_{T2})$. We model the influence of these gate electrodes on the TLF as acting on the detuning bias through $\epsilon = \mathbf{\lambda} \cdot (\mathbf{V} - \mathbf{V}_{0})$, where $\mathbf{\lambda}$ is the electrostatic lever arm vector and $\mathbf{V}_{0}$ is an arbitrary reference voltage in the hyperplane corresponding to the zero detuning point of the TLF. This model captures how device electrostatics perturb the relative energy of the two states of the TLF.

Our phenomenological model for the transition rate of the TLF between state $j$ and $i$ is given by a sum of environment-assisted tunneling and thermally activated rates
\begin{equation}
    \Gamma_{ij}^{\mathrm{tot}} = \Gamma_{ij}^{\mathrm{tunneling}} + \Gamma_{ij}^{\mathrm{activated}},
\end{equation}
where for $\Delta E_{ij} \equiv E_{i} - E_{j}$,
\begin{equation}
    \Gamma_{ij}^{\mathrm{tunneling}} = \vert \langle E_{i} \vert \sigma_{z} \vert E_{j} \rangle \vert^{2} \left[ (1+n(\Delta E_{ji})) J\left( \Delta E_{ji} \right) + n(\Delta E_{ij}) J\left(\Delta E_{ij}\right)\right]
    \label{eq:Tunneling_rate}
\end{equation}
is the Fermi's Golden Rule \cite{phillips1987two} rate of tunneling from eigenstate $\vert E_{j} \rangle$ to $\vert E_{i} \rangle$ of the TLF, $J(x)$ is proportional to the density of bath states at the energy difference $E_{j}-E_{i}$, and $n(\Delta E) = (e^{\Delta E/k_{B}T}-1)^{-1}$ is the Bose-Einstein distribution with $k_{B}$ the Boltzmann constant and $T$ the temperature. 
While this expression can be derived from a spin-boson model \cite{Breuer2002}, it is also consistent with a generalization of the transition rates for electron-assisted tunneling \cite{phillips1987two}.
The detailed balance condition enforces $\Gamma_{ji}^{\mathrm{tunneling}} = \Gamma_{ij}^{\mathrm{tunneling}} e^{-\Delta E_{ji}/k_{B} T}$.
The thermally activated rate term of our model takes the Arrhenius form
\begin{equation}
    \Gamma_{ij}^{\mathrm{activated}} = \gamma_{0} e^{-(E_{b}-\Delta E_{ji}/2)/k_{B}T},
\end{equation}
where $E_{b}$ is a barrier energy and $\gamma_{0}$ is an ``attempt rate'' corresponding to the effective transition rate in the limit of vanishing barrier energy and energy difference $\Delta E_{ij}$. Note that this rate also obeys detailed balance, with $\Gamma_{ji}^{\mathrm{tot}}/\Gamma_{ij}^{\mathrm{tot}} = e^{-\Delta E_{ji}/k_{B} T}$.

While the gate voltage lever arm $\lambda$ captures the control-dependent energy bias between TLF states, the remaining control-dependent quantity we must account for is temperature. The independent parameters governing temperature we consider are (1) the dilution refrigerator mixing chamber temperature $T_{\mathrm{MC}}$ and (2) the single electron transistor (SET) charge sensor conductance, $G_{\mathrm{CS}}$. We assume that $T_{\mathrm{MC}}$ corresponds to the temperature of the lattice far from the TLF, while the charge sensor conductance leads to a local heating effect due to e.g. Joule heating from dissipating a current $I_{\mathrm{SD}} = G_{\mathrm{CS}} V_{\mathrm{SD}}$ across a voltage drop given by the source-drain bias $V_{\mathrm{SD}}$ \cite{Gustafsson2013}. The measured quantity in the experiment is the DAQ signal from RF reflectometry, with an implementation similar to Ref. \cite{connors2020rapid}. Using simulations of a circuit model for our RF reflectometry setup, we find that an approximately linear relationship between DAQ signal $V_{\mathrm{DAQ}}$ and CS conductance $G_{\mathrm{CS}}$ likely holds. This is consistent with what has been observed previously in other RF reflectometry experiments \cite{Reilly2007FastRF}. We note that the relation between DAQ signal and $G_{\mathrm{CS}}$ depends on specifics of device tuning, so we may expect the proportionality $G_{\mathrm{CS}} = \alpha V_{\mathrm{DAQ}}$ to vary somewhat between experiments.

\section{Fitting to the experiment}
\subsection{TLF dipole}
\begin{figure*}[ht!]
\centering
  \includegraphics[width=0.55 \textwidth,trim=0 0 0 0,clip]{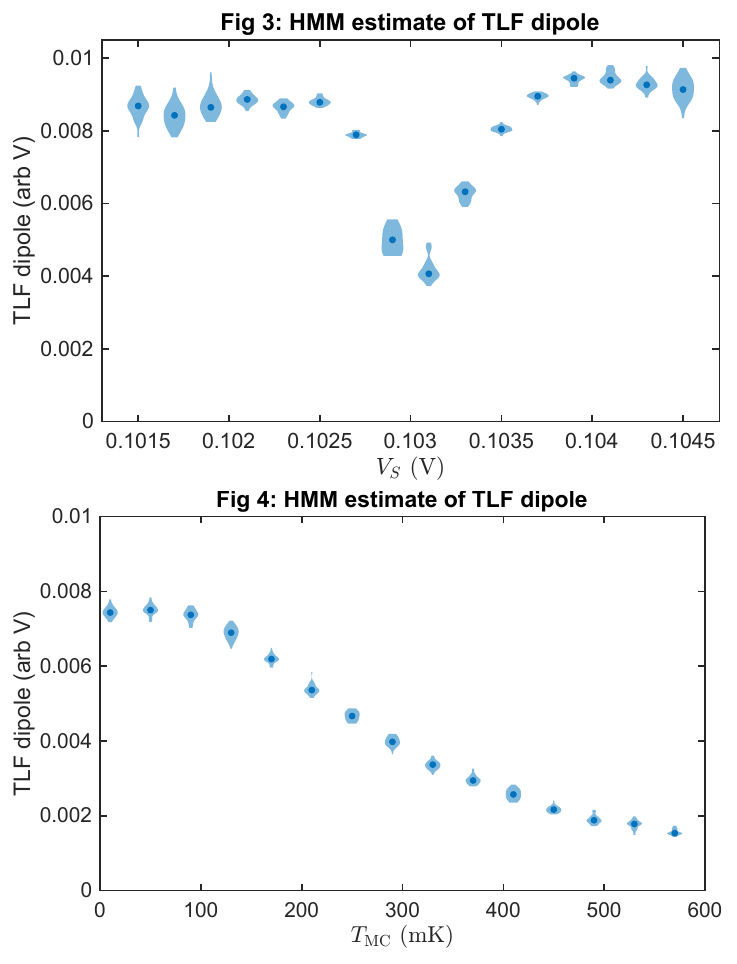}
  \caption{\textbf{Voltage (Temperature) dependence of TLF dipole for Fig 3 (Fig 4) data.} The decrease of observed TLF dipole in the Fig 3 dataset (above) may be an aliasing effect due to the TLF transition rates being comparable to or larger than the 60 Hz sampling rate. However, we expect that the decrease in measured TLF dipole with $T_{\mathrm{MC}}$ in the Fig 4 dataset (below) is due to thermal broadening of the charge sensor peak.}
  \label{fig:Fig23_TLF_dipole}
\end{figure*}

\begin{figure*}[ht!]
\centering
  \includegraphics[width=0.5 \textwidth,trim=0 0 0 0,clip]{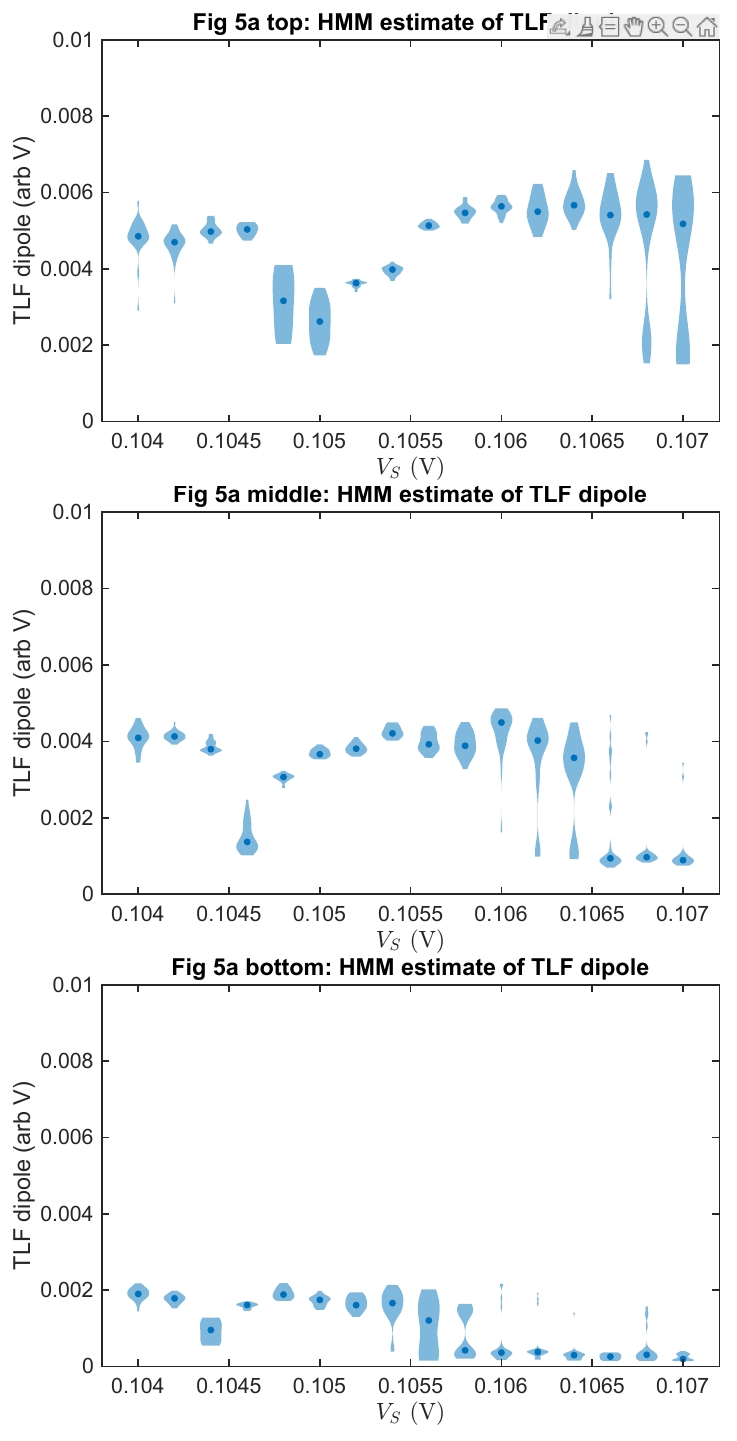}
  \caption{\textbf{Voltage dependence of TLF dipole for Fig 5a data.} As for the Fig. 3 dipoles, we expect the TLF dipole to decrease as a result of aliasing when TLF transition rate becomes comparable to or larger than the 60 Hz sampling rate. The reduction in observed TLF dipole from Fig 5a is consistent with the reduced charge sensitivity when operating the SET at different points on the Coulomb blockade peak.}
  \label{fig:Fig4_TLF_dipole}
\end{figure*}

\begin{figure}[ht!]
\centering
  \includegraphics[width=1 \textwidth,trim=0 0 0 0,clip]{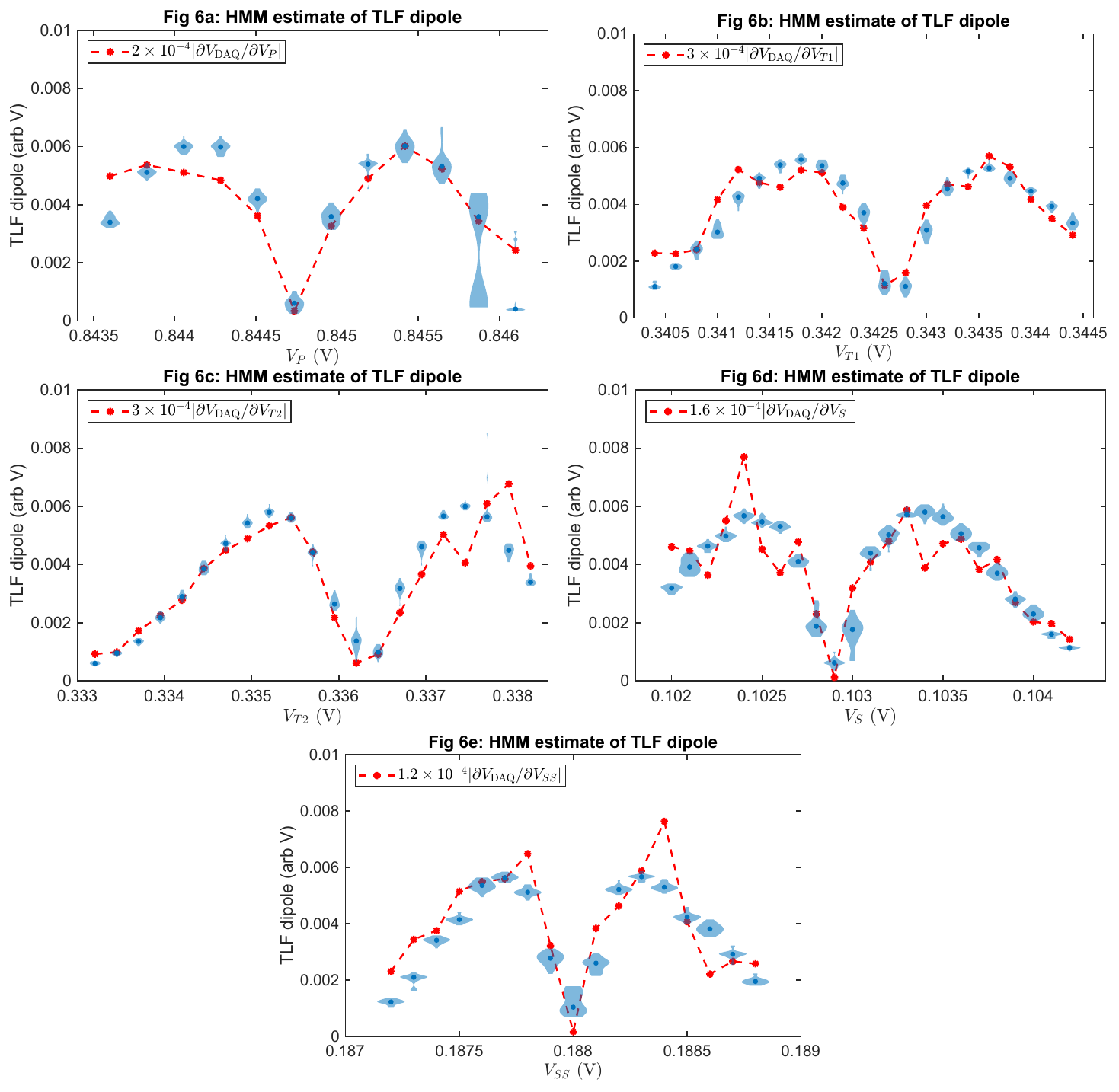}
  \caption{\textbf{Voltage dependence of TLF dipole for Fig 6 data.} The TLF dipole estimated from FHMM fits is plotted in blue. The red curve is proportional to the magnitude of the derivative of the charge sensor signal with respect to the swept gate voltage, suggesting that the TLF electrostatic dipole is fixed and measured dipole variation for these datasets is due to voltage-dependent charge sensitivity.}
  \label{fig:Fig5_TLF_dipole}
\end{figure}

In Figs. \ref{fig:Fig23_TLF_dipole}, \ref{fig:Fig4_TLF_dipole}, and \ref{fig:Fig5_TLF_dipole} we plot the FHMM model-inferred TLF dipole for the various measurement sweeps. We observe significant variation of the dipole that we can explain as artifacts of the measurement rather than intrinsic variation of the electric dipole moment of the TLF. In the sweeps of Figs. 3\&4 and Fig. 5a shown in Figs. \ref{fig:Fig23_TLF_dipole} and \ref{fig:Fig4_TLF_dipole}, the measured TLF dipole appears to drop near the TLF charge transition point where the TLF transition rates are maximal. Since the sampling rate of the measurement is a moving average with sample rate 60 Hz, if the TLF transition rate is a significant fraction of the sample rate we expect aliasing to lead to suppressed TLF dipole. This is indeed what we observe. Note that the change of measured TLF dipole in the Fig. 5a dataset corresponds to the variation of the charge sensor sensitivity as the position on the Coulomb blockade peak is changed. In the dipoles inferred from the Fig. 4 dataset, we observe that the measured TLF dipole decreases with mixing chamber temperature. We believe that this may be due to reduced slope of the Coulomb blockade peak due to thermal broadening.

On the other hand, the variation of measured TLF dipoles of the Fig. 6 datasets shown in Fig. \ref{fig:Fig5_TLF_dipole} is consistent with the slope of the SET Coulomb blockade peak. Using this proportionality, we infer an effective P gate-referred voltage dipole for the TLF of approximately 200 $\mu$V.

\subsection{TLF bias energy}
In order to fit our phenomenological model to the TLF transition rates inferred by the HMM analysis, we first make use of the detailed balance condition $\ln(\Gamma_{10}/\Gamma_{01}) = -\Delta E_{10}/k_{B}T$ and set about fitting $\Delta E_{10}$ and $T$ as a function of the experimentally-adjusted mixing chamber temperature $T_{\mathrm{MC}}$ and gate voltages $\mathbf{V}$. A benefit of first considering the ratio of rates is that dependence on parameters governing the magnitude of the rates themselves drops out, with remaining dependence only on the energy splitting $\Delta E(\mathbf{V},\lambda,t_{c})$ and temperature $T_{\mathrm{TLF}}(\kappa,T_{\mathrm{MC}})$, with the charge sensor conductance signal $V_{\mathrm{DAQ}}$ being itself a function of the gate voltages $\mathbf{V}$.

In our convention for labeling the TLF states 0 and 1, we have
\begin{equation}
    \Delta E_{10} = E_{1} - E_{0} = -\mathrm{sign}(\epsilon) \sqrt{\epsilon^{2} + t_{c}^{2}},
\end{equation}
where for $\epsilon = \lambda \cdot (\mathbf{V} - \mathbf{V}_{0}) < 0$ the low-energy (high-energy) state of the TLF is state 0 (1), respectively. Our model for the temperature dependence is
\begin{equation}
    T_{\mathrm{TLF}} = \left( T_{\mathrm{MC}}^{1+\beta} + \kappa V_{\mathrm{DAQ}}\right)^{1/(1+\beta)},
\end{equation}
where $\kappa$ denotes a charge sensor heating lever arm and we assume $\beta=3$ as the thermal conductivity exponent in all fits. In our model fits, we assume that the effective temperature of the TLF is dictated only by the mixing chamber temperature $T_{\mathrm{MC}}$ and the charge sensor conductance signal $V_{\mathrm{DAQ}}$. That is, we ascribe all ``electron temperature'' contribution to charge sensor heating, since the fits suggest that the significant dynamic range of TLF temperature is correlated strongly with the charge sensor conductance variation. However, it is possible that other sources of heating present even for vanishing SET current, e.g. a higher-temperature nearby two-dimensional electron gas (2DEG), may contribute to an effective TLF temperature elevated above the mixing chamber temperature.

\begin{figure}[ht]
\centering
  \includegraphics[width=0.7 \textwidth,trim=0 0 0 0,clip]{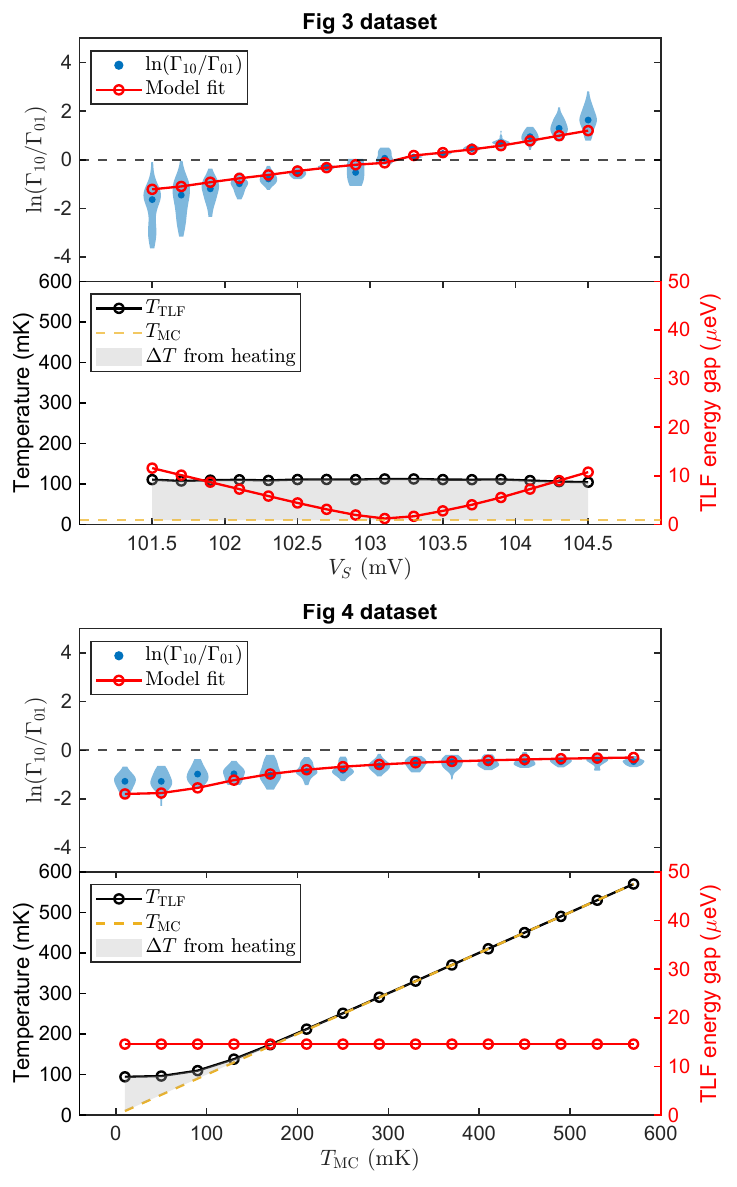}
  \caption{\textbf{Model fit to ratios of TLF transition rates from Fig 3 and Fig 4 data.} (above) Logarithm of ratio of transition rates in blue and model fit in red. (below) Inferred TLF temperature in black, with inferred TLF energy splitting in red. The dashed orange line indicates the mixing chamber temperature.} 
  \label{fig:DetailedBalanceFits_Fig2_3}
\end{figure}

\begin{figure*}[ht]
\centering
  \includegraphics[width=1 \textwidth,trim=5 5 5 5,clip]{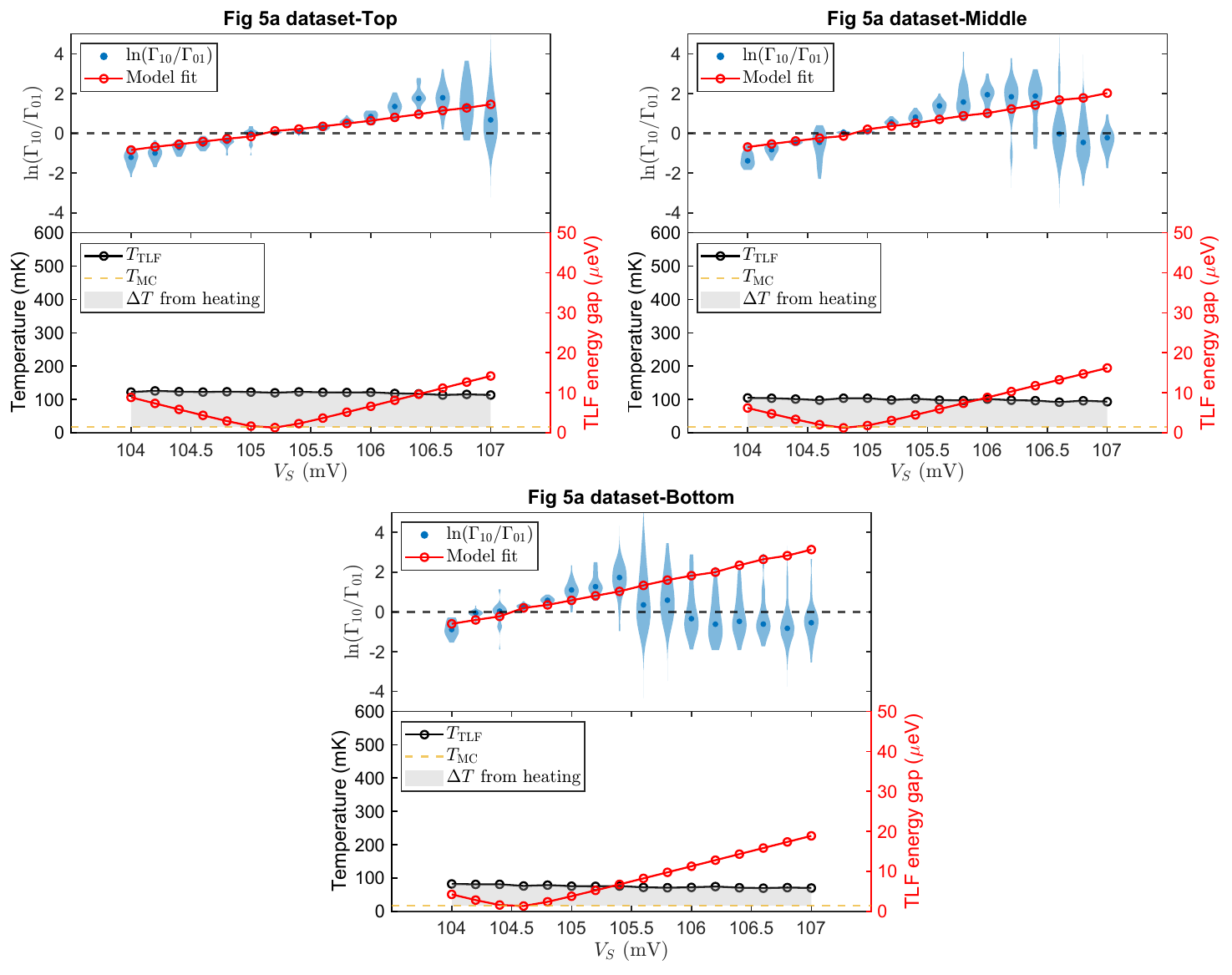}
  \caption{\textbf{Model fit to ratios of TLF transition rates from Fig 5a data.} (above) Logarithm of ratio of transition rates in blue and model fit in red. (below) Inferred TLF temperature in black, with inferred TLF energy splitting in red. The dashed orange line indicates the mixing chamber temperature.}
  \label{fig:DetailedBalanceFits_Fig4}
\end{figure*}

\begin{figure*}[ht]
\centering
  \includegraphics[width=1 \textwidth,trim=5 5 5 5,clip]{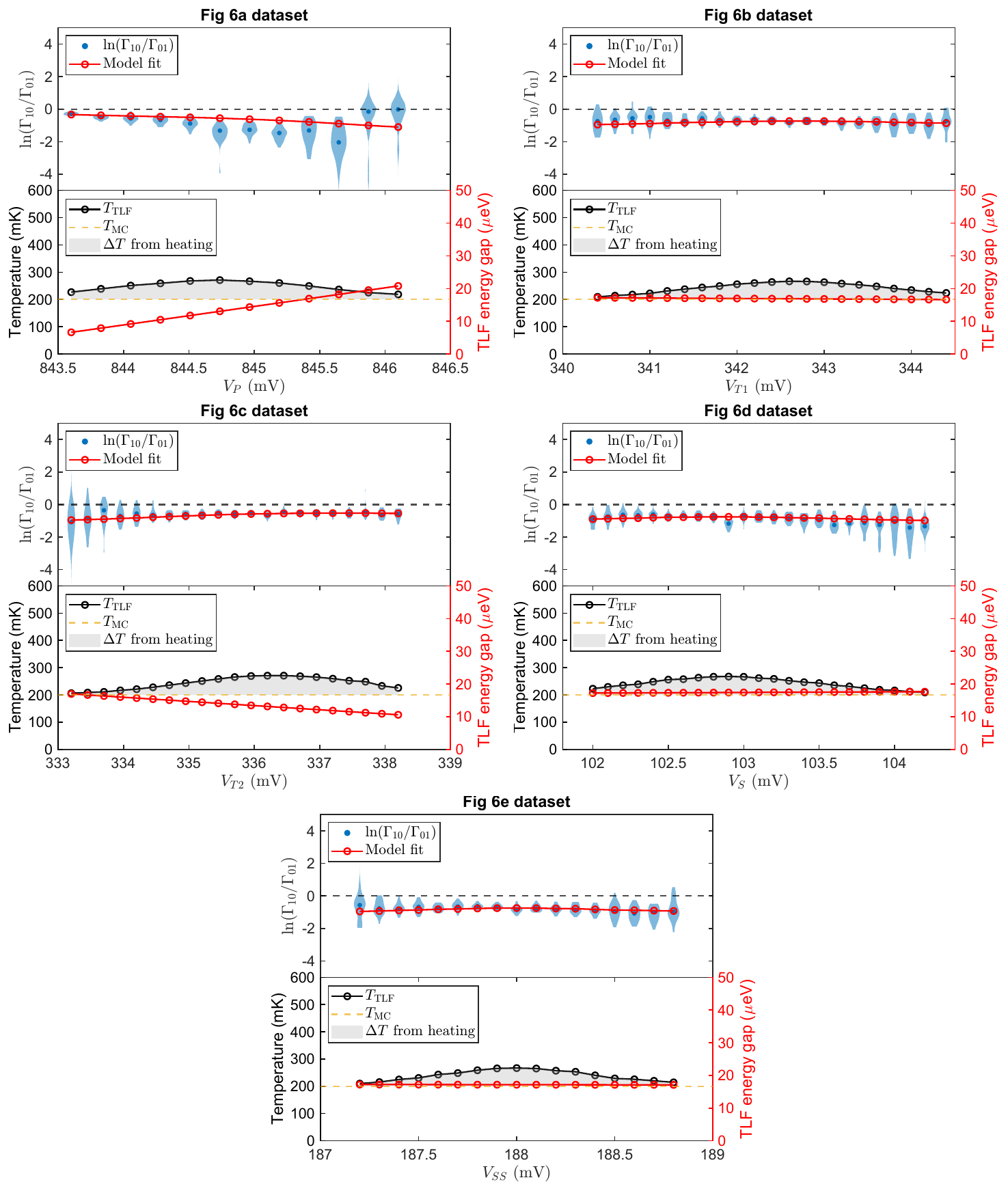}
  \caption{\textbf{Model fit to ratios of TLF transition rates from Figs 6(a-e) data.} (above) Logarithm of ratio of transition rates in blue and model fit in red. (below) Inferred TLF temperature in black, with inferred TLF energy splitting in red. The dashed orange line indicates the mixing chamber temperature.}
  \label{fig:DetailedBalanceFits_Fig5}
\end{figure*}

The model parameters for this fit include the TLF tunnel coupling $t_{c}$, heating lever arm $\kappa$ for each dataset, and lever arm vector $\lambda=(\lambda_{P},\lambda_{S},\lambda_{SS},\lambda_{T1},\lambda_{T2})$. In addition, since the dataset for Fig. 5a was taken several months after the data for Figs. 3, 4, and 6 were taken, we allow for a global bias offset to account for drift and few to tens of mV differences in device tuning voltages.
The fits are shown in Figs. \ref{fig:DetailedBalanceFits_Fig2_3}, \ref{fig:DetailedBalanceFits_Fig4}, and \ref{fig:DetailedBalanceFits_Fig5}. We perform a minimization of the cost function $c(\lambda, t_{c}, \kappa_{\mathrm{Figs3,4}}, \kappa_{\mathrm{Fig5}}, \kappa_{\mathrm{Fig6}}, \Delta \epsilon_{\mathrm{Fig3,4}}, \Delta \epsilon_{\mathrm{Fig5}}, \Delta \epsilon_{\mathrm{Fig6}}) = \sum_{i} (\overline{r}_{i}^{\mathrm{data}} - r_{i}^{\mathrm{model}})^{2}/(\sigma_{i}^{\mathrm{data}})^{2}$
with respect to the aggregate set of measurements shown in Figs.~3, 4, 5a, and 6(a-e), where $\overline{r}_{i}^{\mathrm{data}}$ ($\sigma_{i}^{\mathrm{data}}$) is the mean (standard deviation) of $\ln(\Gamma_{10}/\Gamma_{01})$, respectively, with $i$ indexing each data point. The total number of data points used for the fit in Figs.~\ref{fig:DetailedBalanceFits_Fig2_3}, \ref{fig:DetailedBalanceFits_Fig4}, and \ref{fig:DetailedBalanceFits_Fig5} is 167, with a total of 12 model parameters and cost function value of 153 obtained from a nonlinear minimization using the Nelder-Mead algorithm. We obtain the best-fit values along with 95\% confidence intervals shown in Table \ref{tab:Detailed_balance_fit_params}.  The confidence intervals are calculated by numerically estimating the change in each parameter required to raise $\chi^2(df=12)$ to the value corresponding to the 95\% point of the cumulative distribution~\cite{Avni1976}. Notice that the component of the lever arm $\lambda_{P}$ is by far the largest, consistent with the P gate coupling most strongly to the TLF bias. Finally, we observe a significant variation in the charge sensor heating lever arm between the experimental datasets. This may be due to tuning-dependent changes in the relation between DAQ signal and SET conductance, though further analysis would be needed to confirm this.

\begin{table}[ht!]
    \centering
    % \begin{tabular}{|c|c|c|}
    %      \hline
    %      \textbf{Parameter} & \textbf{Value} & \textbf{Units} \\ 
    %      \hline
    %      Voltage lever arm, $\lambda_{P}$ & -5.7 & $\mu$eV/mV \\
    %      Voltage lever arm, $\lambda_{S}$ & -0.2 & $\mu$eV/mV \\
    %      Voltage lever arm, $\lambda_{SS}$ & 0.1 & $\mu$eV/mV \\
    %      Voltage lever arm, $\lambda_{T1}$ & 0.1 & $\mu$eV/mV \\
    %      Voltage lever arm, $\lambda_{T2}$ & 1.3 & $\mu$eV/mV \\
    %      Tunnel coupling, $t_{c}$ & 1.2 & $\mu$eV \\
    %      CS heating lever arm, $\kappa_{\mathrm{Figs2,3}}$ & 0.004 & $K^{4}/V$ \\
    %      CS heating lever arm, $\kappa_{\mathrm{Fig4}}$ & 0.01 & $K^{4}/V$ \\
    %      CS heating lever arm, $\kappa_{\mathrm{Fig5}}$ & 0.1 & $K^{4}/V$ \\
    %      Voltage bias offset, $\Delta \epsilon_{\mathrm{Fig2,3}}$ & -0.2 & $\mu$eV \\
    %      Voltage bias offset, $\Delta \epsilon_{\mathrm{Fig4}}$ & 0.2 & $\mu$eV \\
    %      Voltage bias offset, $\Delta \epsilon_{\mathrm{Fig5}}$ & -4.0 & $\mu$eV \\
    %      \hline
    % \end{tabular}
    \begin{tabular}{|c|c|c|}
         \hline
         \textbf{Parameter} & \textbf{Value} & \textbf{Units} \\ 
         \hline
         Voltage lever arm, $\lambda_{P}$ & -5.6 [-6.0, -5.0] & $\mu$eV/mV \\
         Voltage lever arm, $\lambda_{S}$ & -0.03 [-0.46, 0.72] & $\mu$eV/mV \\
         Voltage lever arm, $\lambda_{SS}$ & 0.005 [-0.047, 0.053] & $\mu$eV/mV \\
         Voltage lever arm, $\lambda_{T1}$ & 0.24 [-0.36, 1.09] & $\mu$eV/mV \\
         Voltage lever arm, $\lambda_{T2}$ & 1.33 [1.15, 1.60] & $\mu$eV/mV \\
         Tunnel coupling, $t_{c}$ & 1.30 [0.00, 2.90] & $\mu$eV \\
         CS heating lever arm, $\kappa_{\mathrm{Figs3,4}}$ & 0.0041 [0.0014, 0.0185] & $K^{4}/V$ \\
         CS heating lever arm, $\kappa_{\mathrm{Fig5}}$ & 0.011 [0.004, 0.057] & $K^{4}/V$ \\
         CS heating lever arm, $\kappa_{\mathrm{Fig6}}$ & 0.08 [0.02, 0.22] & $K^{4}/V$ \\
         Voltage bias offset, $\Delta \epsilon_{\mathrm{Fig3,4}}$ & -0.16 [-1.66, 1.35] & $\mu$eV \\
         Voltage bias offset, $\Delta \epsilon_{\mathrm{Fig5}}$ & 1.14 [0.14, 2.64]& $\mu$eV \\
         Voltage bias offset, $\Delta \epsilon_{\mathrm{Fig6}}$ & -3.15 [-5.3, -1.1] & $\mu$eV \\
         \hline
    \end{tabular}
    \caption{\textbf{Estimated parameters from fit to ratios of TLF transition rates.} Best-fit values and 95\% confidence intervals for the model parameters constrained by the control-dependent ratios of TLF transition rates $\ln(\Gamma_{10} / \Gamma_{01}) = -\Delta E_{10}/k_{B} T$, where we assume detailed balance to hold. Note that these parameter estimates differ from the full model fit to the individual rates $\Gamma_{01}(\mathbf{V},T)$ and $\Gamma_{10}(\mathbf{V},T)$, though the goodness of fit  is significantly better in this case.}
    \label{tab:Detailed_balance_fit_params}
\end{table}

\subsection{TLF transition rates}
To fit to the full set of transition rates $\lbrace \Gamma_{01},\Gamma_{10}\rbrace$, we define a similar cost function as above using the mean and variances of the corresponding FHMM-inferred transition rates for each repeated measurement point. We consider the sum of tunneling and thermal activation rates, constraining the power law form of the density of states factor $J(\Delta E)$ appearing in the Fermi's Golden Rule expression to agree with the $J(\Delta E) \propto (\Delta E)^{-1}$ dependence determined by fitting to the data shown in Fig. 3c. Again, in these fits we exclude the data points for the transition rates that approach the 60 Hz sampling rate and are expected to exhibit significant aliasing artifacts. To this end, we exclude a data point if $\Gamma_{\mathrm{tot}} = \Gamma_{01} + \Gamma_{10} > 40 \ \mathrm{Hz}$.

For the model fit shown in Fig. 7 of the main text, we obtain a $\chi^{2}$ value of 3349 for a total of 334 data points and 11 model parameters. This goodness of fit is poor, in contrast with the previously discussed fit to the ratio of transition rates. Possible explanations for this include: (1) the model is too simple to capture all details of the parameter dependence, (2) unmodeled confounding parameters such as additional voltage drift or tuning-dependent variation of mapping between charge sensor conductance and DAQ signal, or (3) our nonlinear optimization did not sufficiently explore the parameter space and found only a local optimum.

In our fits to the full dataset, we find that the tunneling component of the transition rate $\Gamma_{ij}^{\mathrm{tunneling}}$ is required to describe the voltage dependence of the transition rates, and the thermal activation component $\Gamma_{ij}^{\mathrm{activated}}$ by itself is not consistent with the data. On the other hand, the data shown in Fig.~6 exhibit a strong dependence on the charge sensor heating effect that, we find, requires the thermal activation component to provide the observed significant dynamic range of transition rates. Further exploration or development of other candidate models for the observed voltage and temperature dependence of the TLF transition rates is warranted.
 %%%% End of modeling section

%\bibliography{noisebib}
